\documentclass[final,1p,times]{elsarticle}
\pdfoutput=1
\usepackage{lineno,hyperref}
\modulolinenumbers[5]

\journal{Astroparticle Physics}
\usepackage[utf8]{inputenc}
\usepackage{cancel}
\usepackage{amssymb,amsmath,mathrsfs,enumerate}
\usepackage{graphicx,rotate,multicol}
\usepackage{float}
\usepackage{subfig}
\usepackage{soul}
\usepackage{mathabx}
\usepackage{eso-pic}
\usepackage{lipsum}

\usepackage{float}
\usepackage{bm}
\usepackage{color}
\usepackage{array}
\usepackage{graphicx}
\usepackage{multirow}
\newcolumntype{P}[1]{>{\centering\arraybackslash}p{#1}}
\newcolumntype{M}[1]{>{\centering\arraybackslash}m{#1}}

\newcommand{\be}{\begin{equation}}
\newcommand{\ee}{\end{equation}}

\newcommand{\een}{\end{subequations}}
\newcommand{\ben}{\begin{subequations}}

\newcommand{\lsim}{\mathrel{\mathop{\kern 0pt \rlap
      {\raise.2ex\hbox{$<$}}}\lower.9ex\hbox{\kern-.190em $ \sim$}}}
\newcommand{\gsim}{\mathrel{\mathop{\kern 0pt
      \rlap{\raise.2ex\hbox{$>$}}}\lower.9ex\hbox{\kern-.190em $\sim$}}}

\definecolor{ourcolor}{rgb}{0.7, 0.25, 0.05}

\let\tilde=\widetilde

\let\bar=\overline

\def \order(#1){{\mathcal O} \left(#1 \right)}

\newcommand{\ba}{\begin{array}}
\newcommand{\ea}{\end{array}}
\newcommand{\bd}{\begin{displaymath}}
\newcommand{\ed}{\end{displaymath}}
\newcommand{\besub}{\begin{subequations}}
\newcommand{\eesub}{\end{subequations}}
\newcommand{\bea}{\begin{eqnarray}}
\newcommand{\eea}{\end{eqnarray}}

\begin{document}

\begin{frontmatter}

\title{Low-mass extension of direct detection bounds on WIMP-quark and WIMP-gluon effective interactions using the Migdal effect}

\author{Gaurav Tomar}
\ead{physics.tomar@tum.de}
\address{Physik-Department, Technische Universit\"at M\"unchen, \protect\\James-Franck-Stra\ss{}e, 85748 Garching, Germany}
\author{Sunghyun Kang}
\ead{francis735@naver.com}
\author{Stefano Scopel}
\ead{scopel@sogang.ac.kr}
\address{Department of Physics, Sogang University, 
Seoul, Korea, 121-742}

\begin{abstract}
Updating a previous analysis where we used elastic nuclear recoils we study the Migdal effect to extend to low WIMP masses the direct detection bounds to operators up to dimension 7 of the relativistic effective field theory describing WIMP interactions with quarks and gluons. To this aim we include in our analysis the data of the XENON1T, SuperCDMS, COSINE-100, and DarkSide-50 experiments and assume a standard Maxwellian for the WIMP velocity distribution. We find that the bounds can reach down to a WIMP mass $\simeq$20 MeV, although in the case of higher--dimension operators the energy scale of the ensuing constraints may be inconsistent with the validity of the effective theory.
\end{abstract}


\end{frontmatter}

\section{Introduction}
\label{sec:introduction}

A worldwide effort has been under way for more than thirty years in the attempt to observe the interactions between the Weakly Interacting Massive Particles (WIMPs) expected to form the Dark Halo of our Galaxy and the nuclear targets of solid--state, liquid, and gaseous detectors appropriately shielded by cosmic rays in Direct Detection (DD) experiments run in underground laboratories around the world~\cite{DD_Schumann2019, Snowmass_Leane2022}. 

The non–observation of new physics at the Large Hadron Collider (LHC) has increasingly constrained the most popular Dark Matter (DM) candidates predicted by extensions of the Standard Model (SM), prompting the need to use bottom--up approaches where the WIMP--nucleus interaction is parameterized in a model--independent way making use of effective models. 
A popular approach is to parameterize the WIMP interaction in terms of the most general relativistically--invariant WIMP--quark and WIMP--gluon operators up to some dimension~\cite{bishara_2017},

\begin{equation}
    {\cal L}_\chi=\sum_q\sum_{a,d}{\cal C}^{(d)}_{a,q}{\cal O}^{(d)}_{a,q}+\sum_{b,d} {\cal C}^{(d)}_b {\cal O}^{(d)}_b,
    \label{eq:eff_L}
\end{equation}
\noindent with ${\cal C}^{(d)}_{a,q}$ and  ${\cal C}^{(d)}_{b}$ some dimensional Wilson coefficients, and $d$ the dimensionality of each operator. 

In particular, in a previous analysis~\cite{relativistic_eft_sogang} we parameterized the Wilson coefficients ${\cal C}$ in terms of an effective scale ${\tilde\Lambda}$,

\begin{equation}
    {\cal C}^{(d)}_{a,q},{\cal C}^{(d)}_{b}=\frac{1}{{\tilde\Lambda}^{d-4}},
    \label{eq:lambda_tilde}
\end{equation}

\noindent and derived the bounds on $\tilde{\Lambda}$ from an extensive list of DD experiments searching for WIMP--nucleus recoils for each of the relativistic operators , ${\cal O}^{(d)}_{a,q}$,  ${\cal O}^{(d)}_b$ up to $d$=7 listed in Eqs.~(\ref{eq:dim5}-\ref{eq:dim7}), assuming a standard Maxwellian velocity distribution for the WIMPs in the halo of our Galaxy.

The bounds derived in the analysis of Ref.~\cite{relativistic_eft_sogang} were restricted to a WIMP mass $m_\chi\gsim$ 600 MeV.
Such loss of sensitivity at low WIMP mass is an intrinsic limitation of nuclear recoil searches, because in such process when $m_\chi$ is too low the minimal amount of energy deposited in the detector by the recoiling nucleus can exceed the experimental threshold of existing experiments only for an incoming WIMP speed larger than the escape velocity in the halo of our Galaxy.

As first pointed out in~\cite{migdal_vergados_1_2005, migdal_vergados_2_2005, migdal_vergados_3_2005, migdal_vergados_4_2005} the production of nuclear excited states triggered by WIMP scattering can produce peculiar experimental signatures and, in particular, the sensitivity of direct detection experiments can be extended to lower WIMP masses~\cite{dama_migdal, Ibe:2017yqa} by making use of the Migdal effect~\cite{migdal_1941} in which the WIMP scattering process triggers the ionization of the recoiling nucleus. This is due to the fact that the ionization or excitation of an electron from an inner orbital can result in extra electronic energy injections, allowing to extend the experimental sensitivity below the threshold for elastic WIMP--nucleus scattering processes. Several experimental collaborations have exploited the Migdal effect in recent studies~\cite{LUX_2018, EDELWEISS_2019, CDEX_2019, XENON_migdal, cosine_migdal, EDELWEISS_2022, supercdms_migdal, ds50_migdal}. The effect has been explored in great details for the standard WIMP interactions with ordinary matter~\cite{Dolan_2017, Essig_2019, Cortona_2020, Flambaum_2020, Knapen_2020, Dey_2020, Bell:2021zkr, Bell_2021_migdal, Wang_2021, Liang_2022, Chatterjee_2022, Angevaare_2022, Cox:2022ekg}, including non-standard WIMP-nucleon interactions using effective field theory (EFT)~\cite{Bell_2019}. However, the analysis presented in~\cite{Bell_2019} is limited as only four non-relativistic operators-$\mathcal{O}_1$, $\mathcal{O}_4$, $\mathcal{O}_6$, $\mathcal{O}_{10}$-have been considered out of 14 different operators possible for spin-$1/2$ DM. Specifically, all the velocity-dependent operators have been left out. In the present work we analysed relativistic effective interactions which, in their non-relativistic limit, extend to non-relativistic operators
 not limited to $\mathcal{O}_1$, $\mathcal{O}_4$, $\mathcal{O}_6$, $\mathcal{O}_{10}$, and contain a more broader picture of the Migdal effect in effective field theories.

The goal of the present letter is to use the Migdal effect to extend to $m_\chi\lsim$ 600 MeV the bounds discussed in Ref.~\cite{relativistic_eft_sogang} on the effective operators of Eq.~(\ref{eq:eff_L}). To this aim we will
consider the dedicated analyses on the Migdal effect from four direct detection experiments: XENON1T~\cite{XENON_migdal}, COSINE-100~\cite{cosine_migdal}, SuperCDMS~\cite{supercdms_migdal}, and DarkSide-50~\cite{ds50_migdal}.
Our main results are given in the exclusion plots of Figs.~\ref{fig:dim-5}, \ref{fig:dim-6} and \ref{fig:dim-7_1}.

The paper is organised as follows. In Section~\ref{sec:expected_rates} we outline the procedure to calculate the expected rate for Migdal events triggered by WIMP--nucleus scattering; in Section~\ref{sec:rel_eft} we summarize the effective models considered in the present analysis and already studied in Ref.~\cite{relativistic_eft_sogang}. Finally, we provide our quantitative results in Section~\ref{sec:analysis} and our conclusions in Section~\ref{sec:conclusion}.
\section{Expected rates}
\label{sec:expected_rates}
In the Migdal effect the WIMP--nucleus scattering process is accompanied by the ejection of an electron from the recoiling nucleus with the
ensuing deposit of an electromagnetic (EM) signal in the detector. As a consequence, the techniques developed to discriminate between nuclear recoil processes and EM energy depositions are not applied, since the latter, rather than being only due to background, are produced also by WIMPs. 
This requires to re--analyze the experimental data and allows to lower the threshold compared to the analyses that look for elastic recoils. In particular, the ionisation event rate in an experiment due to the Migdal effect is calculated as~\cite{Ibe:2017yqa},

\begin{equation}
\label{diff_rate_migdal0}
 \frac{dR}{dE_{det}}=\int_0^{\infty} d E_R \int_{v_{min}(E_R)}^\infty dv_T \frac{d^3 R_{\chi T}}{dE_R dv_T dE_{det}}
\end{equation}

\noindent with
\begin{equation}
 \label{diff_rate_migdal}
 \frac{d^3R_{\chi T}}{dE_RdE_{det}dv_T}=\frac{d^2R_{\chi T}}{dE_R dv_T}\times \frac{1}{2\pi}\sum_{n,l}\frac{d}{dE_e}p^c_{q_e}(nl\rightarrow(E_e)).
 \end{equation}
 
\noindent The Migdal effect is usually negligible compared to the standard signal from elastic nuclear recoils, unless for very low WIMP masses for which the latter is below threshold. At higher WIMP masses a transition between Migdal and nuclear recoil takes place, until eventually the latter dominates.  For the WIMP masses, target nuclei and the intervals for detected energy, $E_{det}$ considered in the analysis of Section~\ref{sec:analysis},  we have verified that the EM energy produced by the recoil of the nucleus during the Migdal process remains negligible~\footnote{One exception is sodium, for which the effect of nuclear recoil is non-negligible in the interval of $E_{det}$ used by COSINE-100 if a standard value found in the literature for the quenching factor, $q_{Na}=0.3$, is used~\cite{Bernabei:2003za, Coarasa:2018qzs}. However, in our analysis we use for sodium the quenching factor measured by COSINE-100~\cite{COSINE-100:2019brm}, that is significantly smaller. In this case the effect of nuclear recoil can be neglected.  }  In this case, neglecting the contribution from the nuclear recoil, $E_{det}\simeq E_{EM}$, where the total injected EM energy, $E_{EM}$ is the sum of the outgoing electron energy, $E_e$ and of the energy from de-excitation $E_{nl}$, where $n$ and $l$ are the initial quantum numbers of the ionized electron~\footnote{For this reason in the following $E_{det}$ will be identified with $E_{EM}$.}. Moreover, in Eqs.~(\ref{diff_rate_migdal0}, \ref{diff_rate_migdal}) $E_R$ is the nuclear recoil energy,  $v_T$ is the WIMP speed in the reference frame of the nuclear center of mass and,
\begin{equation}
v_{\rm min}(E_R)=\frac{m_T E_R+\mu_T E_{EM}}{\mu_T\sqrt{2m_T E_R}}.
\label{eq:vmin}
\end{equation}
In the above equation, $m_T$ is the nucleus mass and $\mu_T$ represents the reduced mass. Finally $p^c_{q_e}$ represents the ionisation probability,  while $q_e=m_e\sqrt{2E_R/m_T}$ is the average momentum transfer to an individual electron in the rest frame of the target nucleus. 

In particular in our calculation we obtain the differential scattering rate spectrum $\frac{d^2R_{\chi T}}{dE_R dv_T}$ using the WimPyDD~\cite{wimpydd} code, and utilise the ionization probabilities $p^c_{qe}$ calculated in Ref.~\cite{Ibe:2017yqa}. 

In our analysis we have followed the approach of XENON1T, DarkSide-50, SuperCDMS, and COSINE-100 adopting the isolated atom approximation, which is valid for inner shell electrons, neglecting the contribution from valence electrons. In particular, for solid-state detectors, as pointed out in~\cite{cosine_migdal, supercdms_migdal} most of the valence shell electrons contribution to the signal is below the experimental threshold, so that neglecting it implies a negligible decrease in the expected rate.

\section{Relativistic effective models}
\label{sec:rel_eft}

In this Section we outline the procedure that we follow to obtain the numerical results of Section~\ref{sec:analysis}. For definiteness, in our analysis we closely follow for the effective operators the notations of~\cite{bishara_2017, directdm}, used also in~\cite{relativistic_eft_sogang}.
\begin{figure}
\begin{center}
  \includegraphics[width=0.49\textwidth]{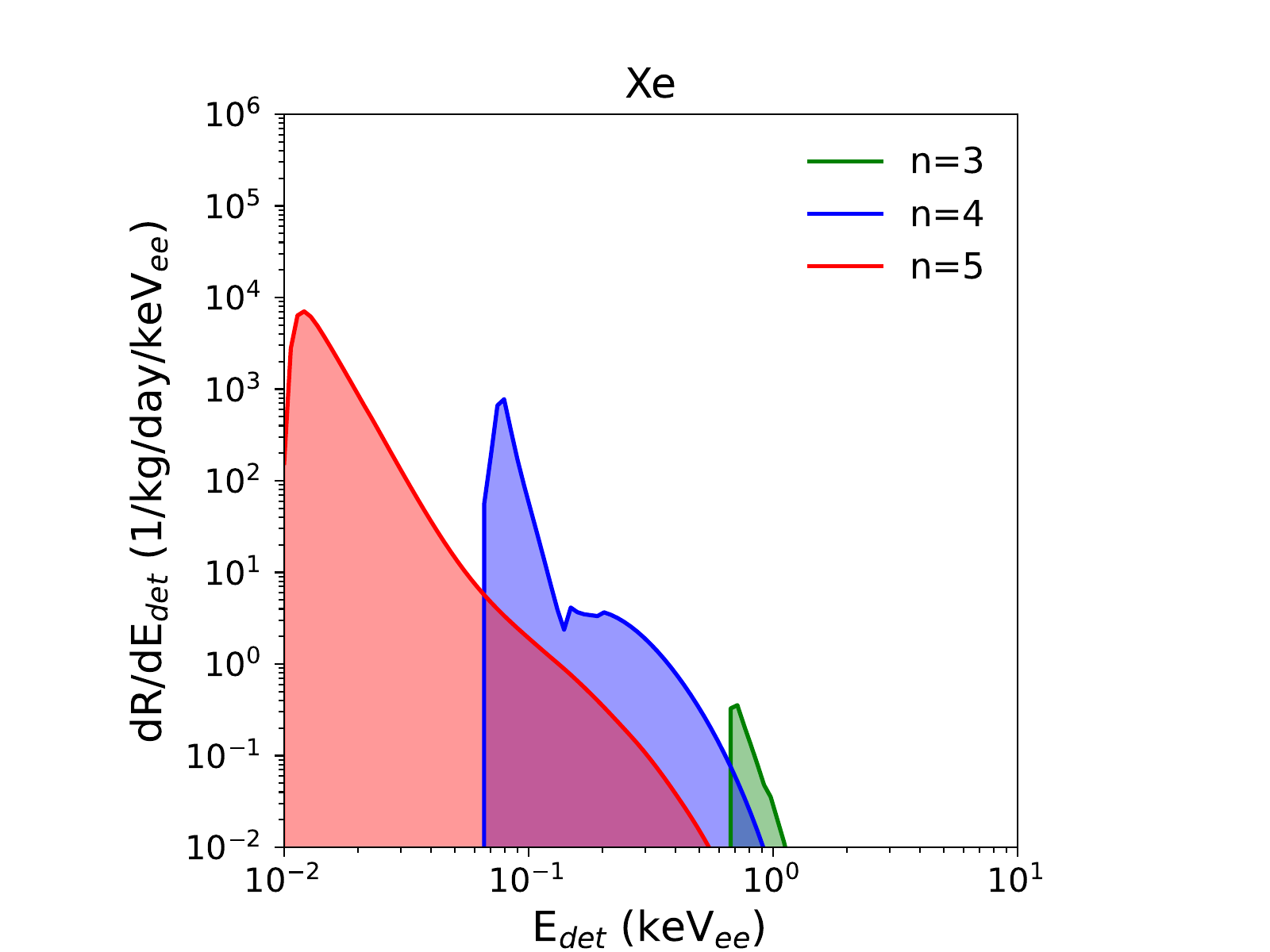}
  \includegraphics[width=0.49\textwidth]{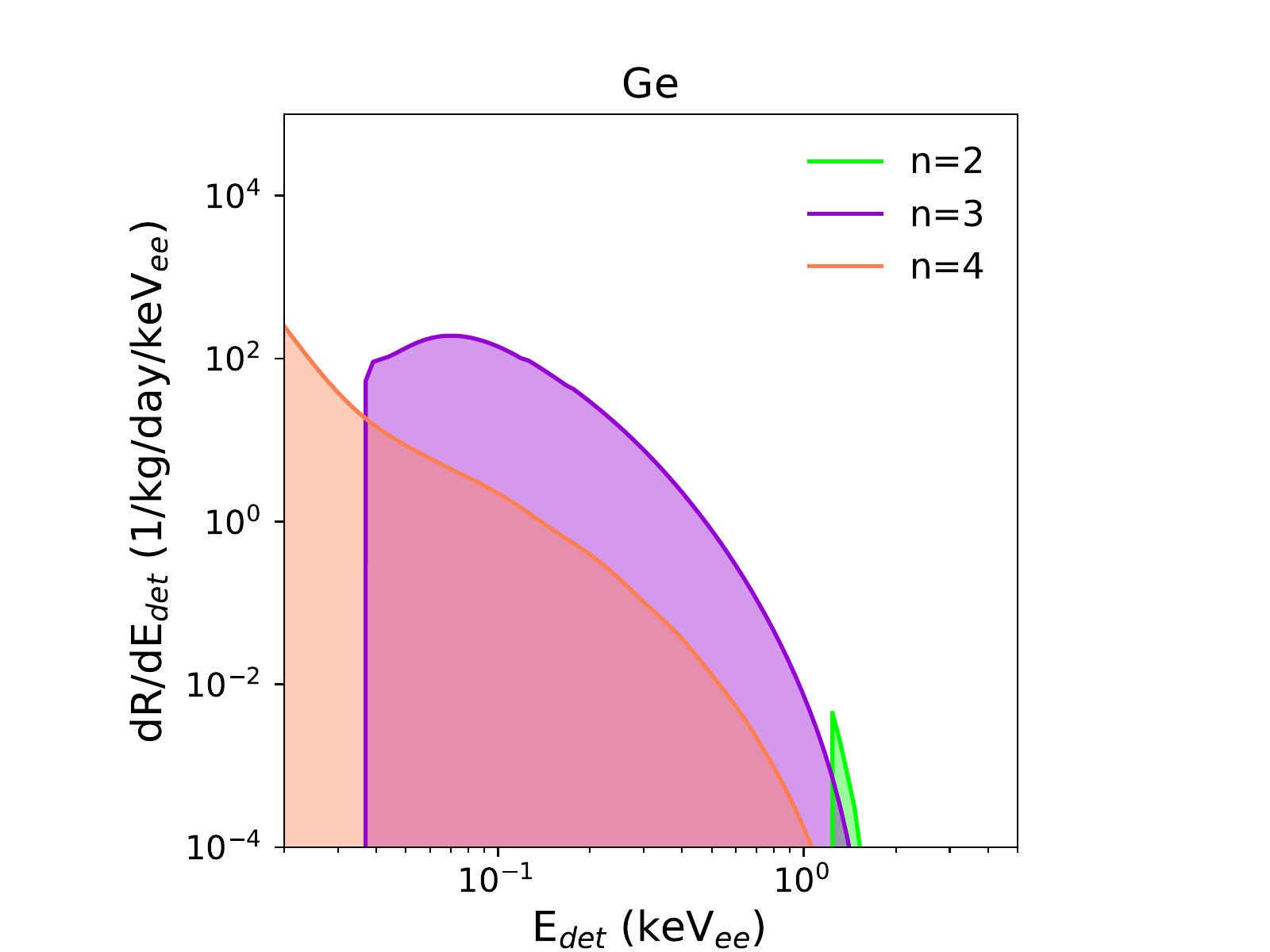}\\
\end{center}
\begin{center}
  \includegraphics[width=0.49\textwidth]{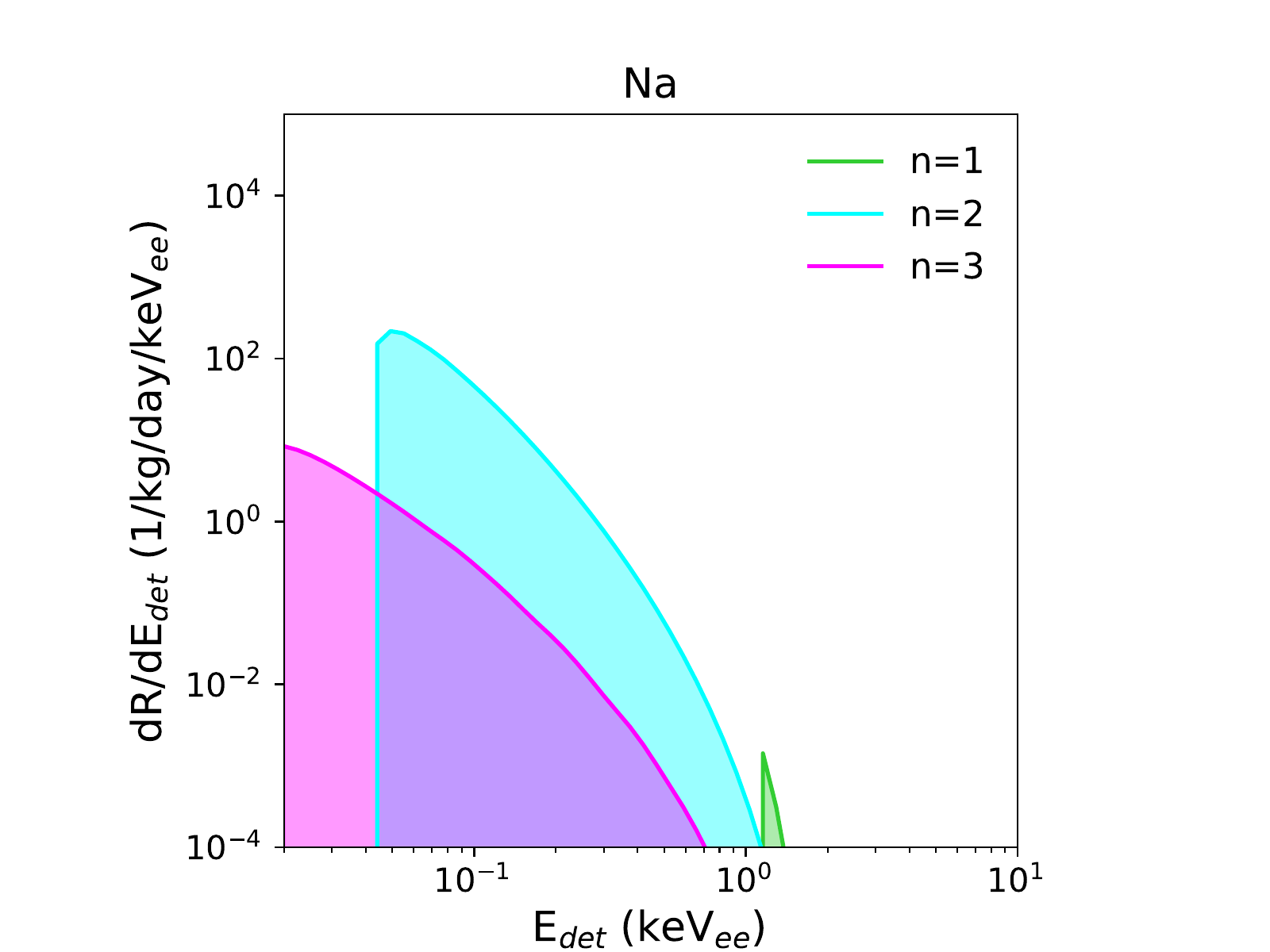}
  \includegraphics[width=0.49\textwidth]{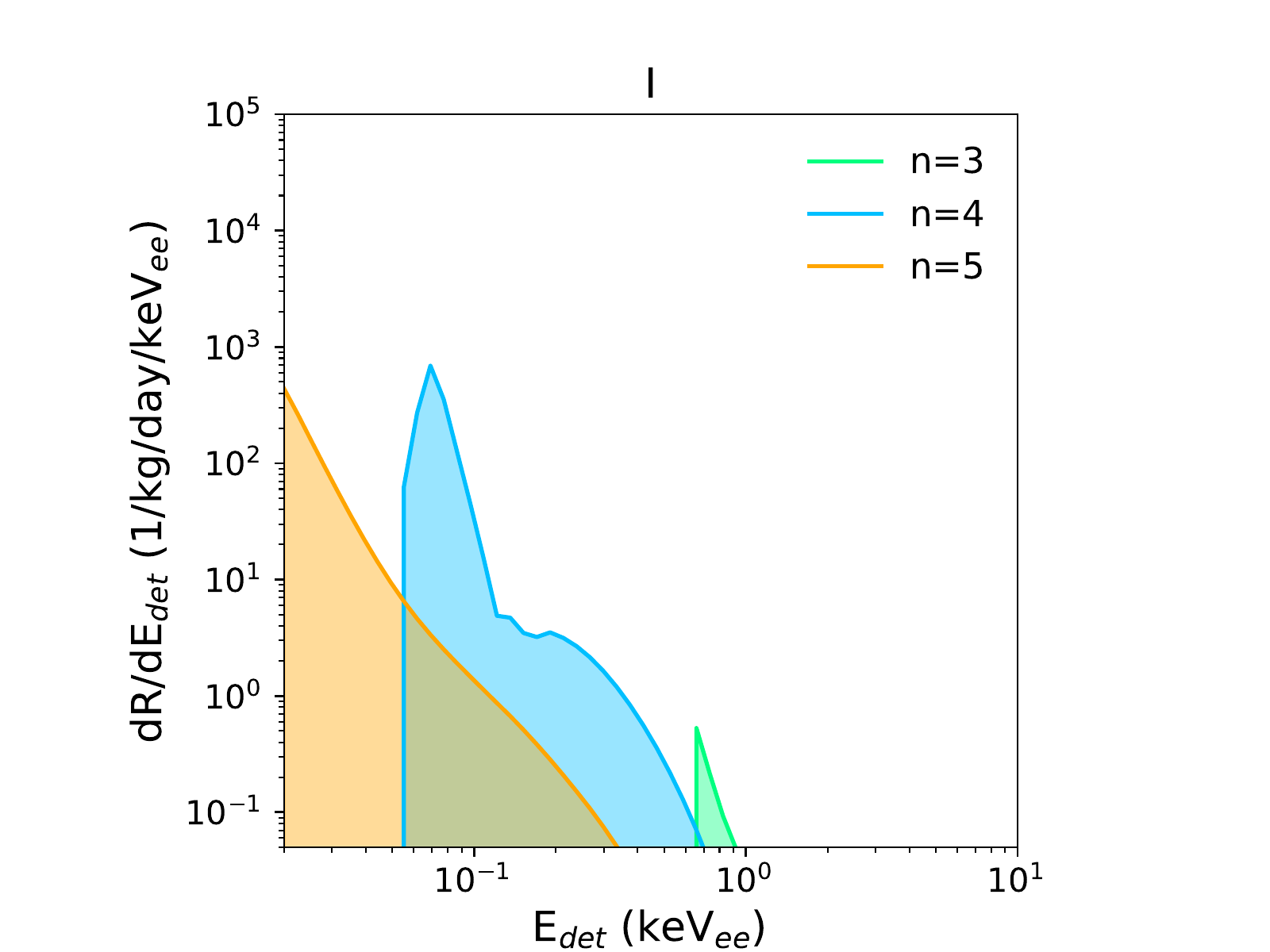}\\
\end{center}
\begin{center}
  \includegraphics[width=0.49\textwidth]{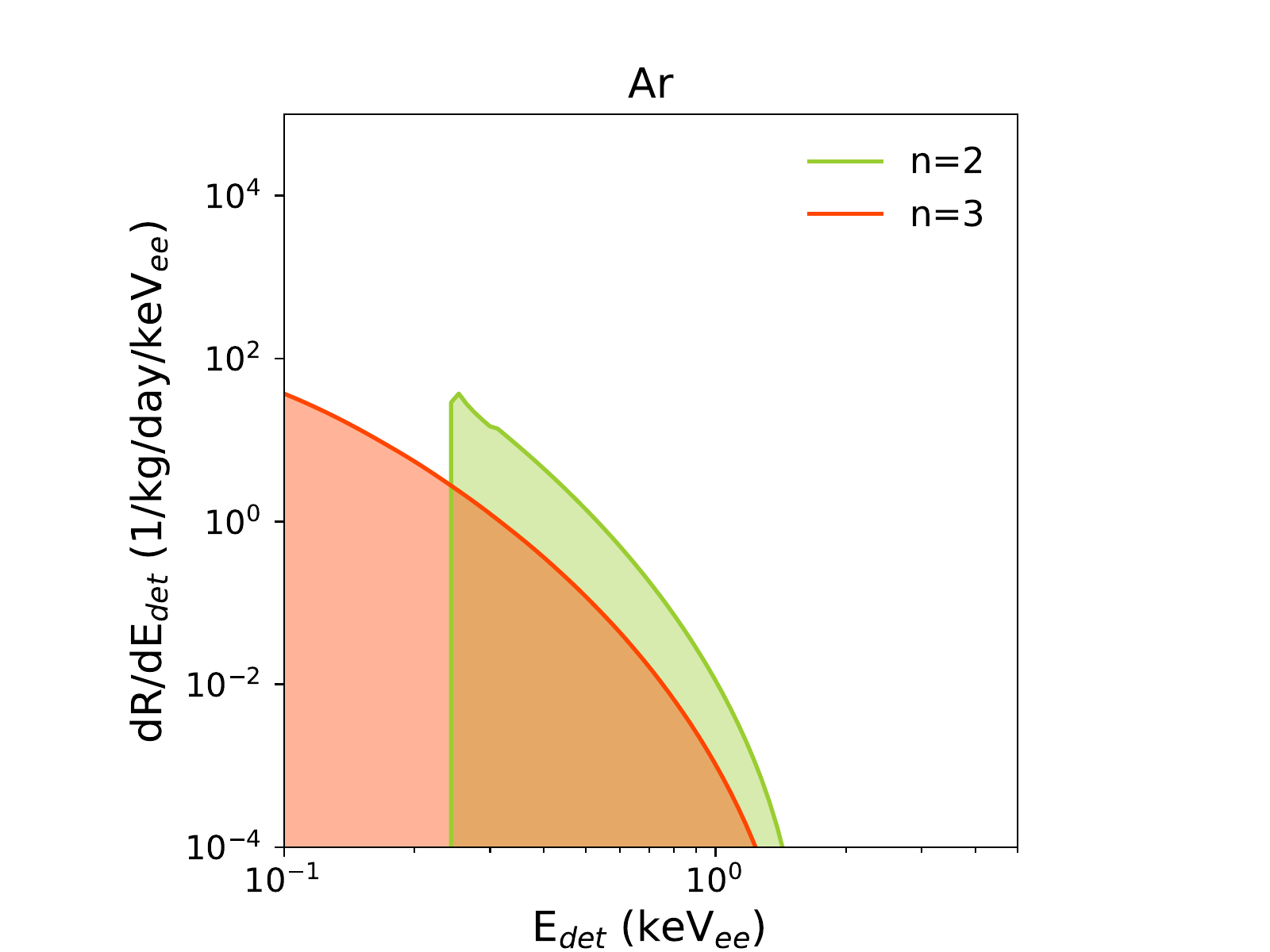}
\end{center}
\caption{The Migdal differential rate for $m_\chi=0.5$ GeV, taking $\tilde\Lambda=1$ GeV and magnetic dipolar interaction ${\cal Q}_{1}^{(5)}$, with Xe, Ge, Na, I, and Ar targets.  In the horizontal axis the energy $E_{det}$ includes the contribution to the deposited EM energy from nuclear recoils, which is only relevant above the threshold for elastic scattering events. When scattering events are below such threshold their contribution is negligible and $E_{det}\simeq E_{EM}$. The different color shadings correspond to the ionisation rates from $n=1,2,3,4$, and $5$ depending upon the considered targets.\label{fig:diff_rate}}
\end{figure}
In particular, we consider the two dimension-five operators,
\begin{equation}
\label{eq:dim5}
{\cal Q}_{1}^{(5)} = \frac{e}{8 \pi^2} (\bar \chi \sigma^{\mu\nu}\chi)
 F_{\mu\nu} \,, \qquad {\cal Q}_2^{(5)} = \frac{e }{8 \pi^2} (\bar
\chi \sigma^{\mu\nu} i\gamma_5 \chi) F_{\mu\nu} \,,
\end{equation}
where $F_{\mu\nu}$ is the electromagnetic field strength tensor and
$\chi$ is the DM field, assumed here to be a Dirac particle. Such
operators correspond, respectively, to magnetic--dipole and
electric--dipole DM and imply a long--range
interaction~\cite{delnobile_2018}.  The dimension-six
operators are,
\begin{eqnarray}
{\cal Q}_{1,q}^{(6)} & =& (\bar \chi \gamma_\mu \chi) (\bar q \gamma^\mu q)\,,
 {\cal Q}_{2,q}^{(6)} = (\bar \chi\gamma_\mu\gamma_5 \chi)(\bar q \gamma^\mu q)\,, \nonumber
  \\ 
{\cal Q}_{3,q}^{(6)} & =& (\bar \chi \gamma_\mu \chi)(\bar q \gamma^\mu \gamma_5 q)\,,
   {\cal Q}_{4,q}^{(6)} = (\bar
\chi\gamma_\mu\gamma_5 \chi)(\bar q \gamma^\mu \gamma_5 q)\,.\label{eq:dim6}
\end{eqnarray}
In our analysis we also consider the following dimension-seven operators,
\begin{eqnarray}
{\cal Q}_1^{(7)} & =& \frac{\alpha_s}{12\pi} (\bar \chi \chi)
 G^{a\mu\nu}G_{\mu\nu}^a\,, 
  {\cal Q}_2^{(7)} = \frac{\alpha_s}{12\pi} (\bar \chi i\gamma_5 \chi) G^{a\mu\nu}G_{\mu\nu}^a\,,\nonumber
 \\
{\cal Q}_3^{(7)} & =& \frac{\alpha_s}{8\pi} (\bar \chi \chi) G^{a\mu\nu}\widetilde
 G_{\mu\nu}^a\,, 
 {\cal Q}_4^{(7)} = \frac{\alpha_s}{8\pi}
(\bar \chi i \gamma_5 \chi) G^{a\mu\nu}\widetilde G_{\mu\nu}^a \,, \nonumber
\\
{\cal Q}_{5,q}^{(7)} & =& m_q (\bar \chi \chi)( \bar q q)\,, 
{\cal
  Q}_{6,q}^{(7)} = m_q (\bar \chi i \gamma_5 \chi)( \bar q q)\,,\nonumber
  \\
{\cal Q}_{7,q}^{(7)} &=& m_q (\bar \chi \chi) (\bar q i \gamma_5 q)\,, 
{\cal Q}_{8,q}^{(7)}  = m_q (\bar \chi i \gamma_5 \chi)(\bar q i \gamma_5
q)\,, \nonumber  
 \\
{\cal Q}_{9,q}^{(7)} & =& m_q (\bar \chi \sigma^{\mu\nu} \chi) (\bar q \sigma_{\mu\nu} q)\,, 
{\cal Q}_{10,q}^{(7)}  = m_q (\bar \chi  i \sigma^{\mu\nu} \gamma_5 \chi)(\bar q \sigma_{\mu\nu}
q)\,. \label{eq:dim7} 
\end{eqnarray}
\noindent Here, $q=u,d,s$ denote the light quarks,
$G_{\mu\nu}^a$ is the QCD field strength tensor, while $\widetilde
G_{\mu\nu} = \frac{1}{2}\varepsilon_{\mu\nu\rho\sigma} G^{\rho\sigma}$
is its dual, and $a=1,\dots,8$ are the adjoint color indices. For all the operators of Eqs.(\ref{eq:dim5}--\ref{eq:dim7}) we assume flavor conservation .

The detailed expression for the calculation of the differential rate $\frac{d^2R_{\chi T}}{dE_R dv_T}$ in Eq.~(\ref{diff_rate_migdal}) is provided in Section 2 of \cite{sogang_scaling_law_nr}, which has been implemented in the WimPyDD code~\cite{wimpydd}. In particular, in the non-relativistic limit the differential cross
section is proportional to the squared amplitude,

\be
\frac{d\sigma_T}{d E_R}=\frac{2 m_T}{4\pi v_T^2}\left [ \frac{1}{2 j_{\chi}+1} \frac{1}{2 j_{T}+1}|\mathcal{M}_T|^2 \right ],
\label{eq:dsigma_de}
\ee

\noindent with $m_T$ the nuclear mass, $j_T$, $j_\chi$ the spins of the target nucleus and of the WIMP, where $j_\chi=1/2$, and \cite{haxton2},

\begin{equation}
  \frac{1}{2 j_{\chi}+1} \frac{1}{2 j_{T}+1}|\mathcal{M}_T|^2=
  \frac{4\pi}{2 j_{T}+1} \sum_{\tau=0,1}\sum_{\tau^{\prime}=0,1}\sum_{k} R_k^{\tau\tau^{\prime}}\left [c^{\tau}_i,c_j^{\tau^{\prime}},(v^{\perp}_T)^2,\frac{q^2}{m_N^2}\right ] W_{T k}^{\tau\tau^{\prime}}(y).
\label{eq:squared_amplitude}
\end{equation}

\noindent In the above expression the squared amplitude
$|\mathcal{M}_T|^2$ is summed over initial and final spins, the
$R_k^{\tau\tau^{\prime}}$'s are WIMP response functions which depend
on the couplings $c^{\tau}_j$ as well as the transferred momentum
$\vec{q}$, while,

\begin{equation}
(v^{\perp}_T)^2=v^2_T-v_{min}^2,
\label{eq:v_perp}
\end{equation}
where,
\begin{equation}
v_{min}^2=\frac{q^2}{4 \mu_{T}^2}=\frac{m_T E_R}{2 \mu_{T}^2},
\label{eq:vmin1}
\end{equation}

\noindent represents the minimal incoming WIMP speed required to
impart the nuclear recoil energy $E_R$. Moreover, in Eq.
(\ref{eq:squared_amplitude}) the $W^{\tau\tau^{\prime}}_{T k}(y)$'s
are nuclear response functions and the index $k$ represents different
effective nuclear operators, which, under the assumption that the
nuclear ground state is an approximate eigenstate of $P$ and $CP$, can
be at most eight: following the notation in \cite{haxton1,haxton2},
$k$=$M$, $\Phi^{\prime\prime}$, $\Phi^{\prime\prime}M$,
$\tilde{\Phi}^{\prime}$, $\Sigma^{\prime\prime}$, $\Sigma^{\prime}$,
$\Delta$, $\Delta\Sigma^{\prime}$. The $W^{\tau\tau^{\prime}}_{T
  k}(y)$'s are function of $y\equiv (qb/2)^2$, where $b$ is the size
of the nucleus. For the target nuclei $T$ used in most direct
detection experiments the functions $W^{\tau\tau^{\prime}}_{T k}(y)$,
calculated using nuclear shell models, have been provided in
Refs.~\cite{haxton2,catena}. Details about the definitions of both the
functions $R_k^{\tau\tau^{\prime}}$'s and $W^{\tau\tau^{\prime}}_{T
  k}(y)$'s can be found in \cite{haxton2}.
In our analysis for the WIMP local density we take $\rho_{loc}$=0.3
GeV/cm$^3$ and for the velocity distribution we assume a standard
isotropic Maxwellian with velocity dispersion 220 km/s, truncated at an escape velocity of 550 km/s. 
\begin{figure}
\begin{center}
  \includegraphics[width=0.49\textwidth]{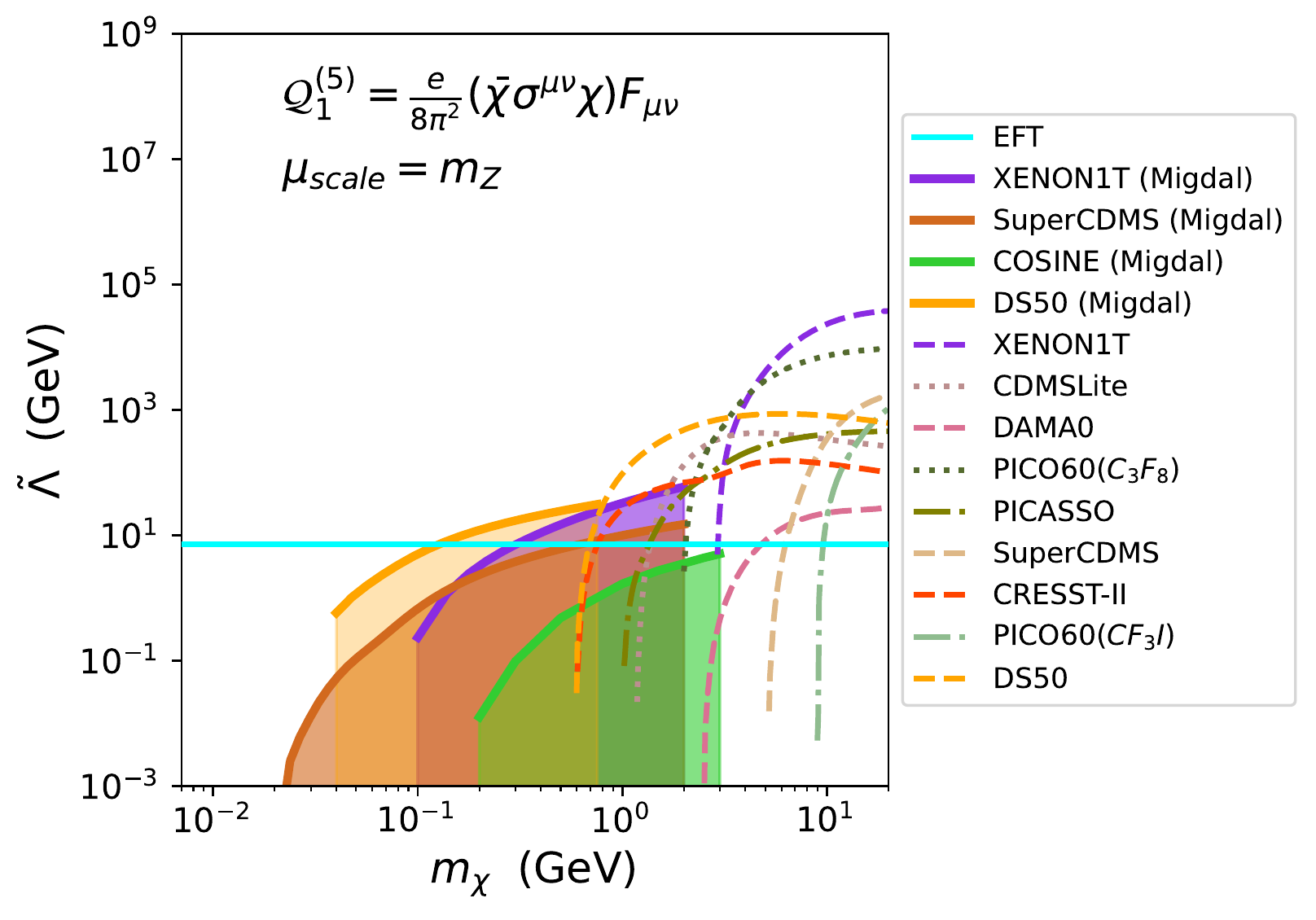}
  \includegraphics[width=0.49\textwidth]{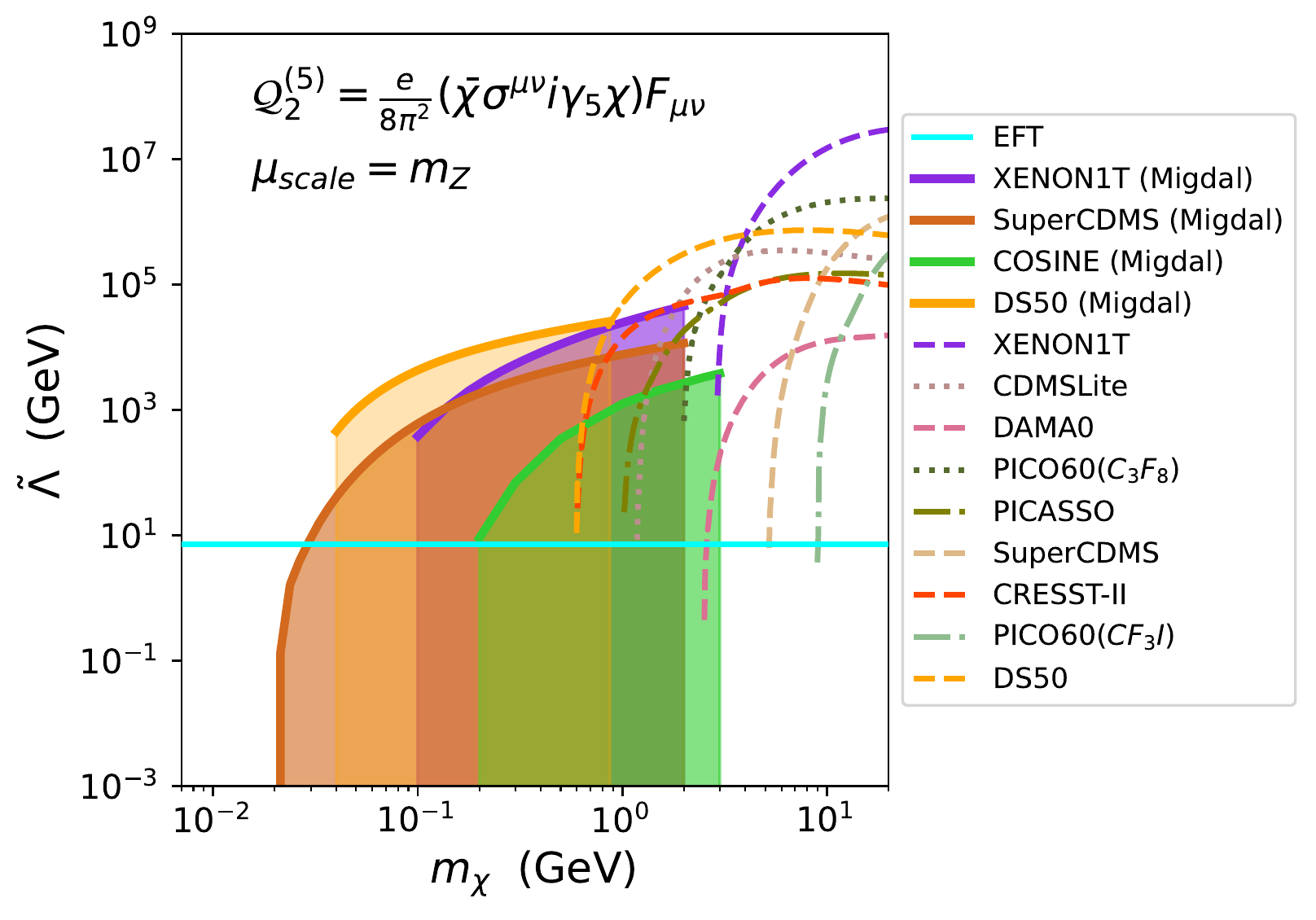}\\
\end{center}
\caption{Lower bound on the effective scale $\tilde\Lambda$ for the operators ${\cal Q}_{1,q}^{(5)}$ (left) and ${\cal Q}_{2,q}^{(5)}$ (right). We fixed the dimensional couplings $C^{(5)}_{1,q}$ and $C^{(5)}_{2,q}$ at the EW scale $\mu_{scale}=m_Z$. In the region below the solid cyan
line the limits are inconsistent with the validity of the EFT as described in Section~\ref{sec:analysis}.\label{fig:dim-5}}
\end{figure}
\section{Analysis}
\label{sec:analysis}
The expected event rate of the Migdal effect of Eq.~(\ref{diff_rate_migdal})  is given by the product of the WIMP-nucleus scattering rate $\frac{d^2R_{\chi T}}{dE_R dv_T}$ and of the ionization probability $p^c_{q_e}$.
\begin{figure}[ht]
\begin{center}
  \includegraphics[width=0.49\textwidth]{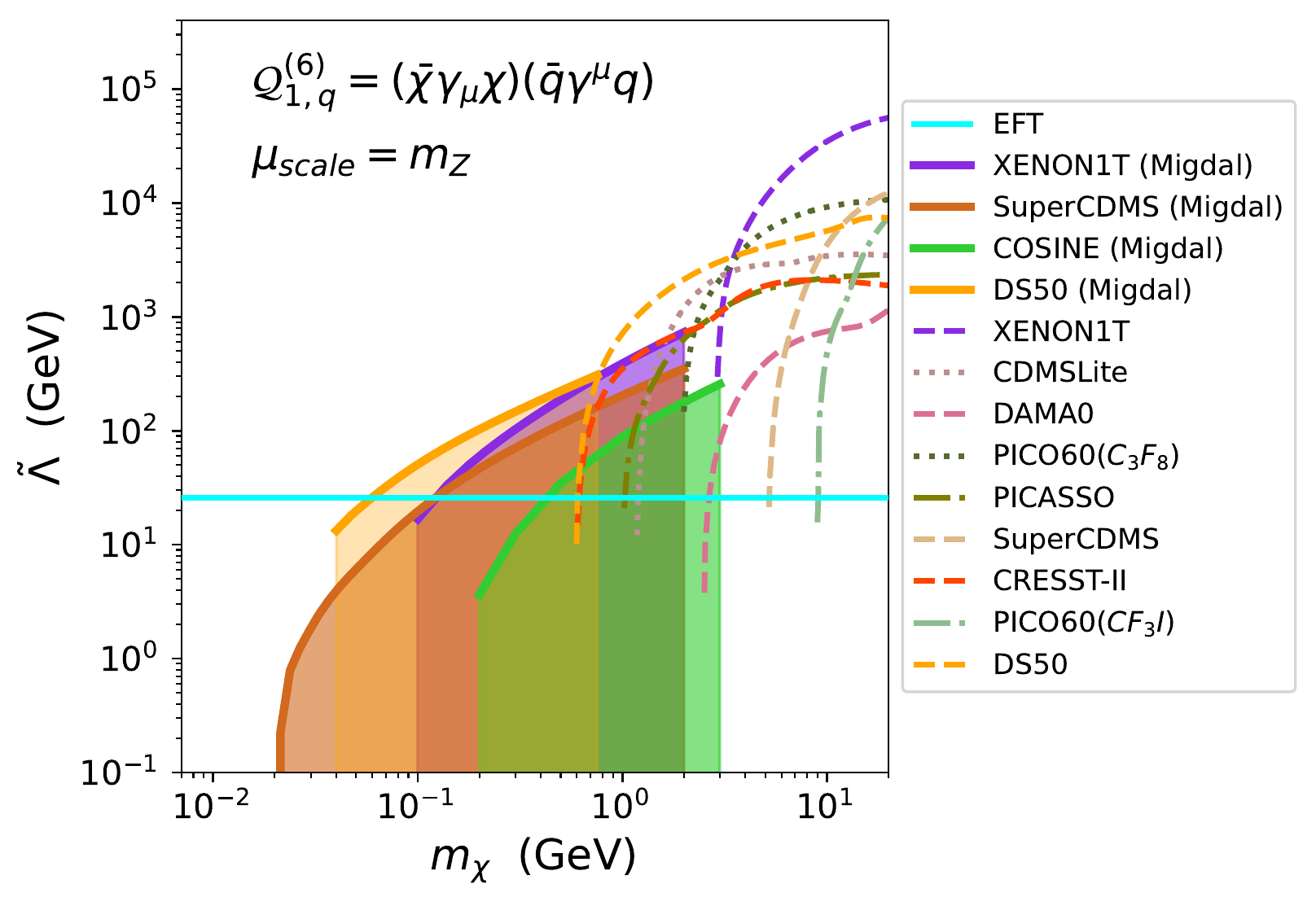}
  \includegraphics[width=0.49\textwidth]{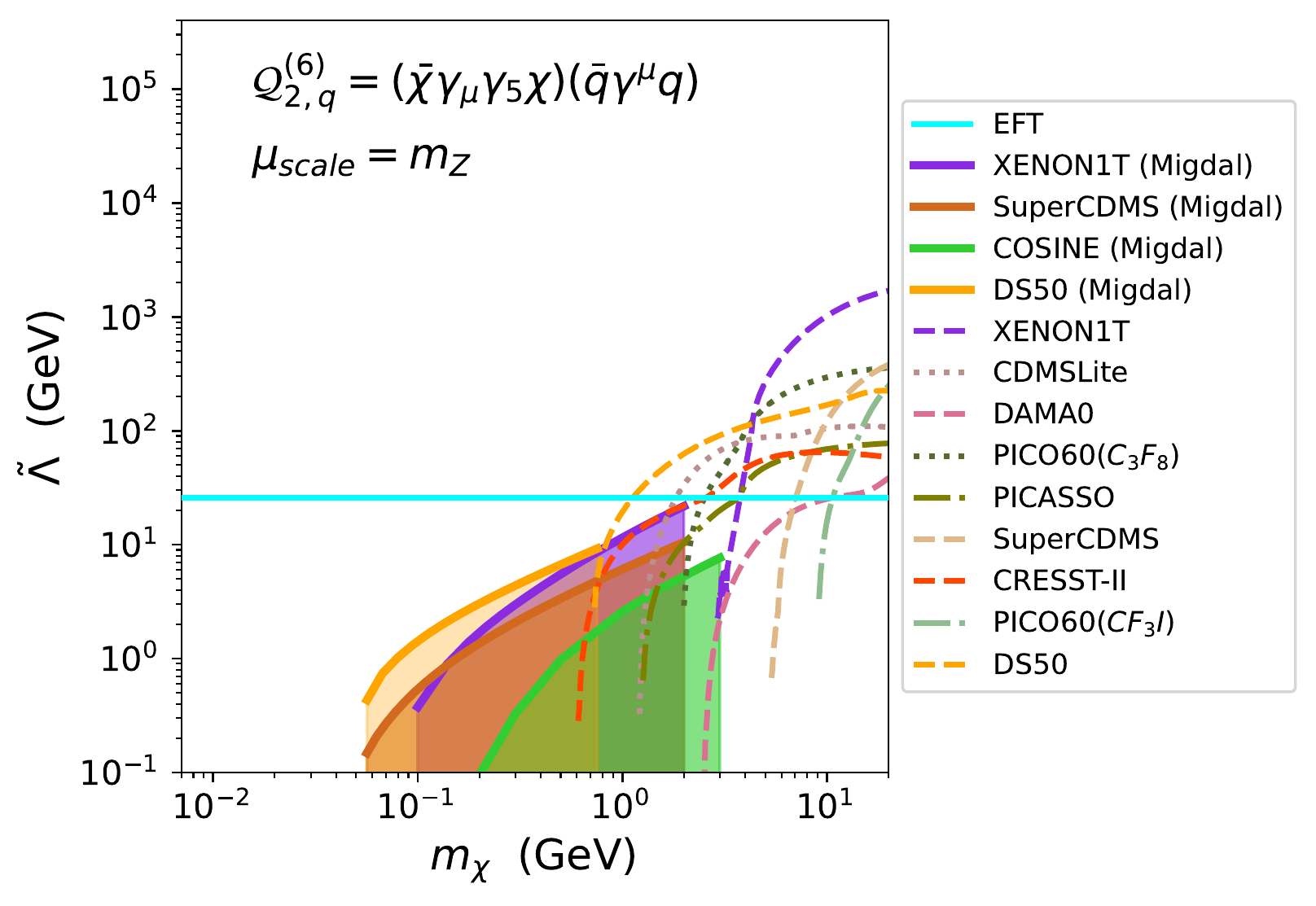}\\
  \includegraphics[width=0.49\textwidth]{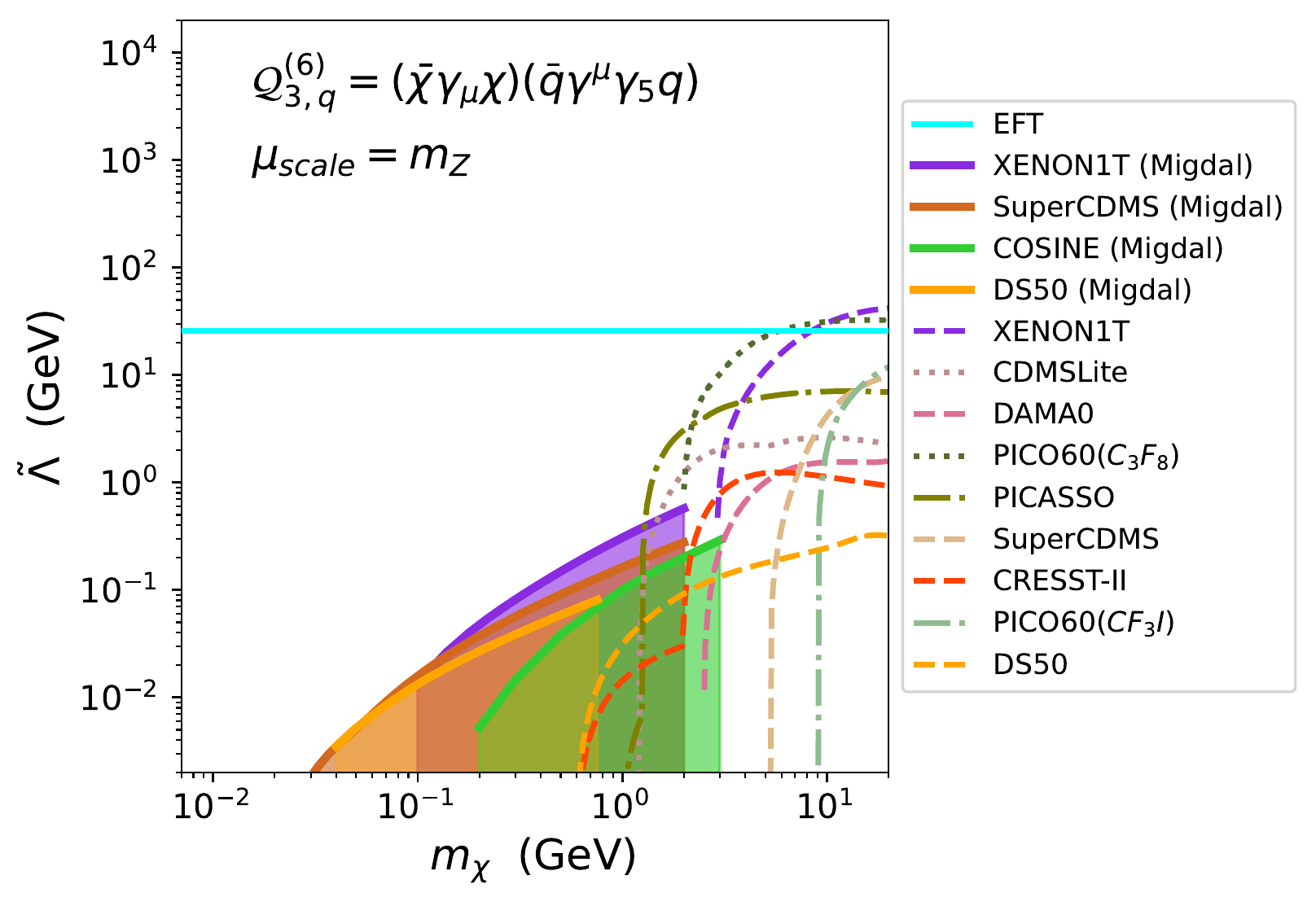}
  \includegraphics[width=0.49\textwidth]{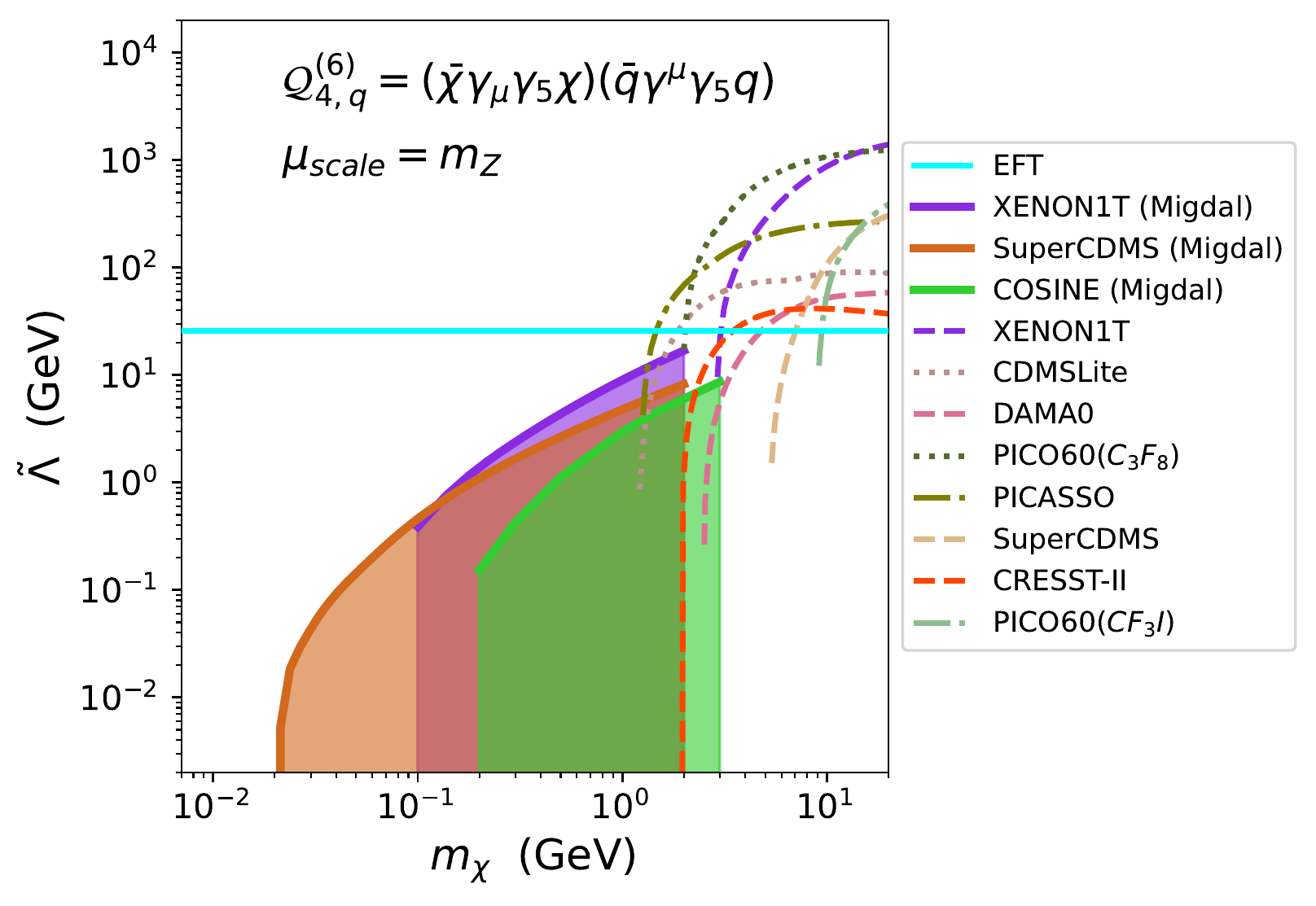}\\
\end{center}
\caption{Same as Fig.~\ref{fig:dim-5} for ${\cal Q}_{1,q}^{(6)}$ (top-left), ${\cal Q}_{2,q}^{(6)}$ (top-right), ${\cal Q}_{3,q}^{(6)}$ (bottom-left), and ${\cal Q}_{4,q}^{(6)}$ (bottom-right).\label{fig:dim-6}}
\end{figure}
As shown in Eq.~(\ref{eq:vmin}) due to the electron emission the kinematics of the scattering rate $R_{\chi T}$ is modified compared to the elastic case. In particular Eq.~(\ref{eq:vmin}) describes the same kinematics of inelastic DM~\cite{inelastic}, where a low--mass DM eigenstate $\chi$ upscatters to a higher--mass state $\chi^{\prime}$ with mass splitting $\delta$ = $m_{\chi^{\prime}}-m_\chi$. In the case of the Migdal effect the energy $E_{EM}$ of the emitted electron takes the place of $\delta$. The calculation of $\frac{d^2R_{\chi T}}{dE_R dv_T}$ can be handled in a straightforward way by the \verb|wimp_dd_rate| routine of WimPyDD~\cite{wimpydd}, that includes the argument \verb|delta| for inelastic scattering. In particular we fixed \verb|delta| equal to $E_{EM}\simeq E_e$, and integrated the scattering rate over the full range of the undetected nuclear scattering energy $E_R$. 

Moreover, in our analysis, we fixed the dimensional couplings ${\cal C}_{a,q}^{(d)}$, ${\cal C}_{b}^{(d)}$ at the Electroweak (EW) scale $\mu_{scale}=m_Z$ and used the DirectDM~\cite{directdm} code to obtain the Wilson coefficients at the WIMP-nucleon interaction scale. The Wilson coefficients obtained in this way were then used in WimPyDD~\cite{wimpydd} for the calculation of WIMP-nucleus scattering rate. As for the ionisation probabilities $p^c_{q_e}$ we adopted those provided in~\cite{Ibe:2017yqa} for Xenon, Sodium, Iodine, Germanium, and Argon, corresponding to the targets of XENON1T~\cite{XENON_migdal}, COSINE-100~\cite{cosine_migdal}, SuperCDMS~\cite{supercdms_migdal}, and DarkSide-50~\cite{ds50_migdal}.

In Fig.~\ref{fig:diff_rate} we provide one explicit example of the differential event rate for the Migdal effect in the case of a magnetic dipole interaction (${\cal Q}^{(5)}_1$) off Xe, Ge, Na, I and Ar targets. The different color shadings correspond to the ionisation rates for $n=1,2,3,4,5$ shells. It is worth mentioning that available DD experiments are sensitive to only some of the shells (partially or fully) that contribute to the Migdal event rate, due to their energy threshold.

\begin{figure}[ht]
\begin{center}
  \includegraphics[width=0.49\textwidth]{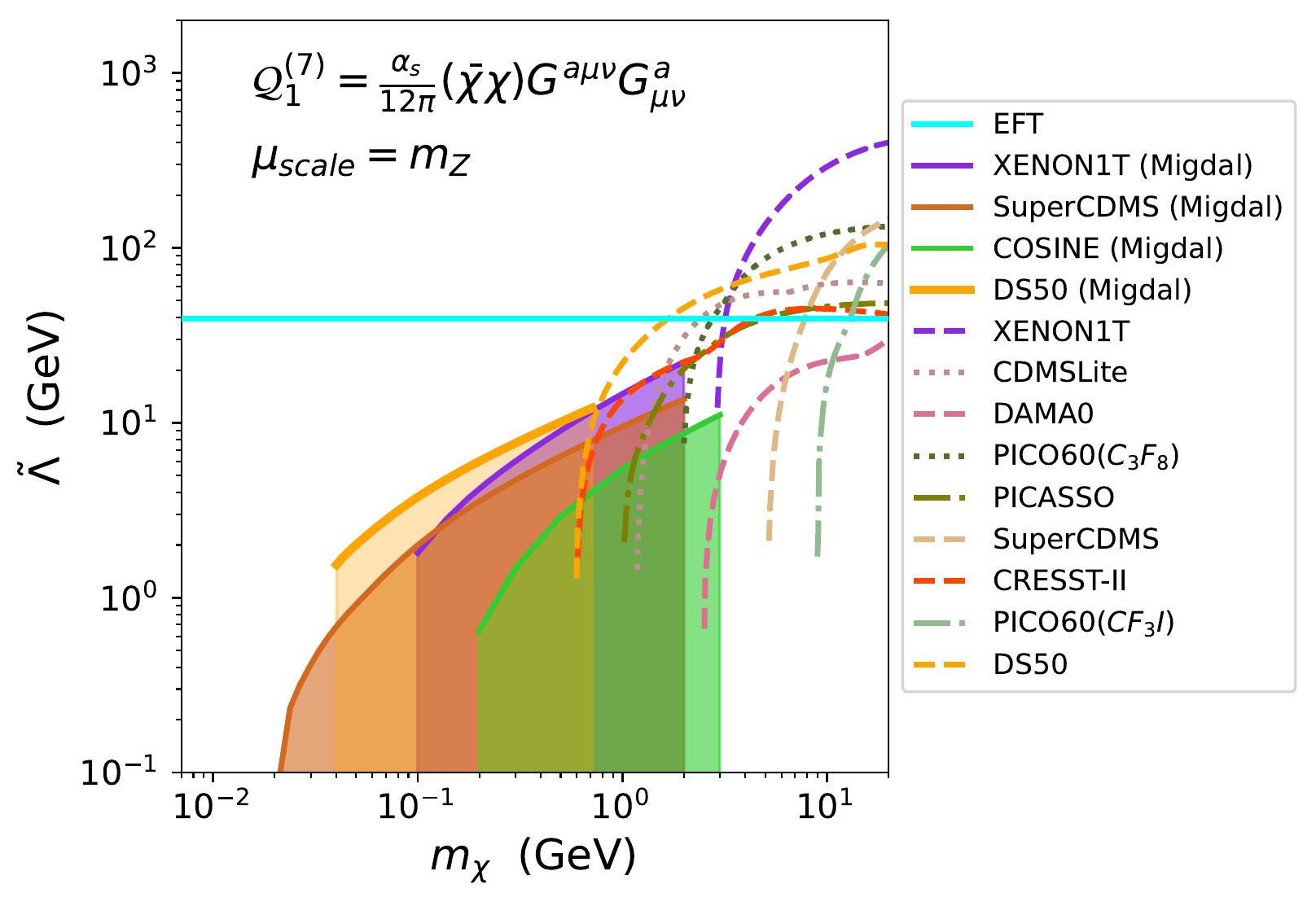}
  \includegraphics[width=0.49\textwidth]{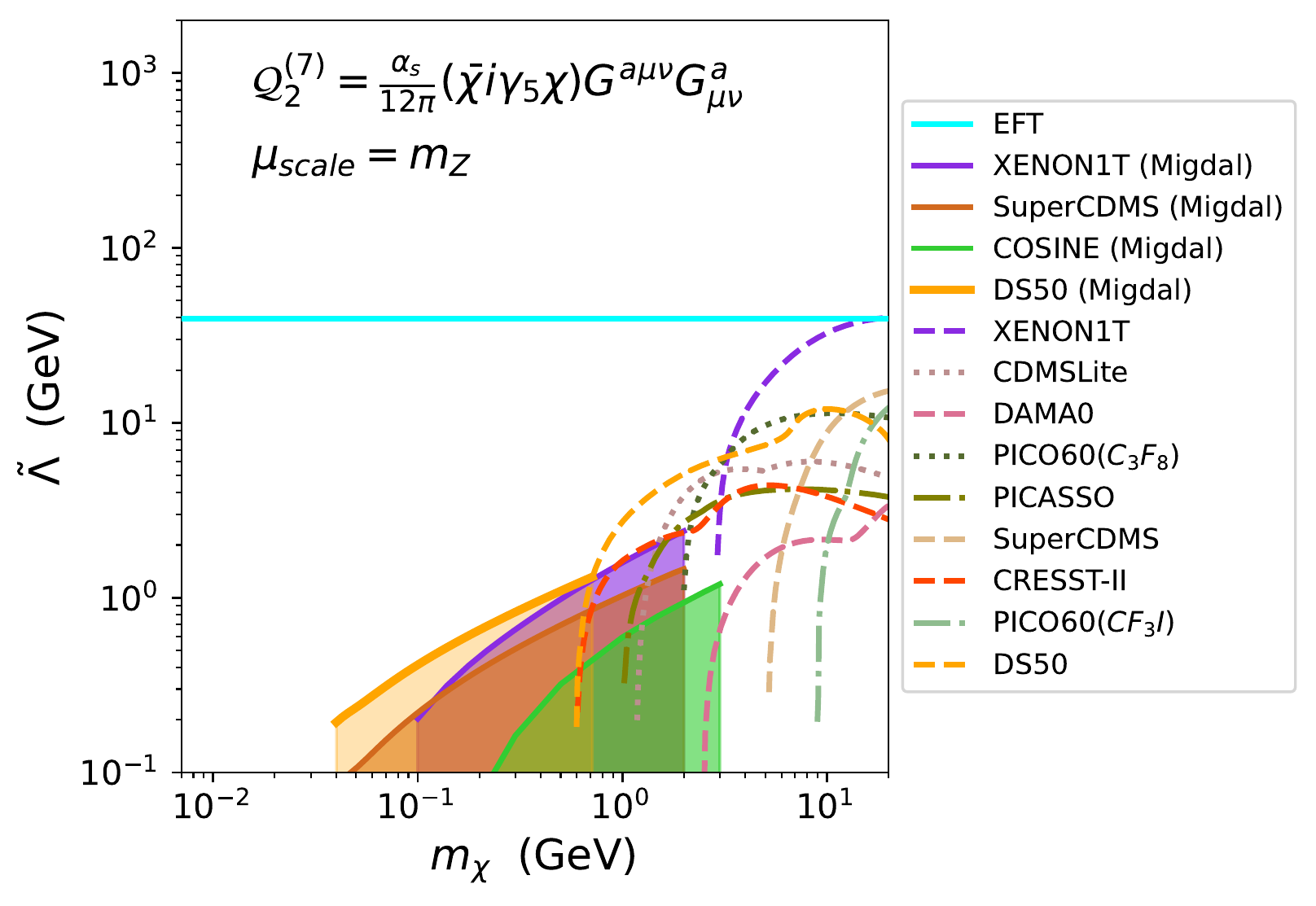}\\
  \includegraphics[width=0.49\textwidth]{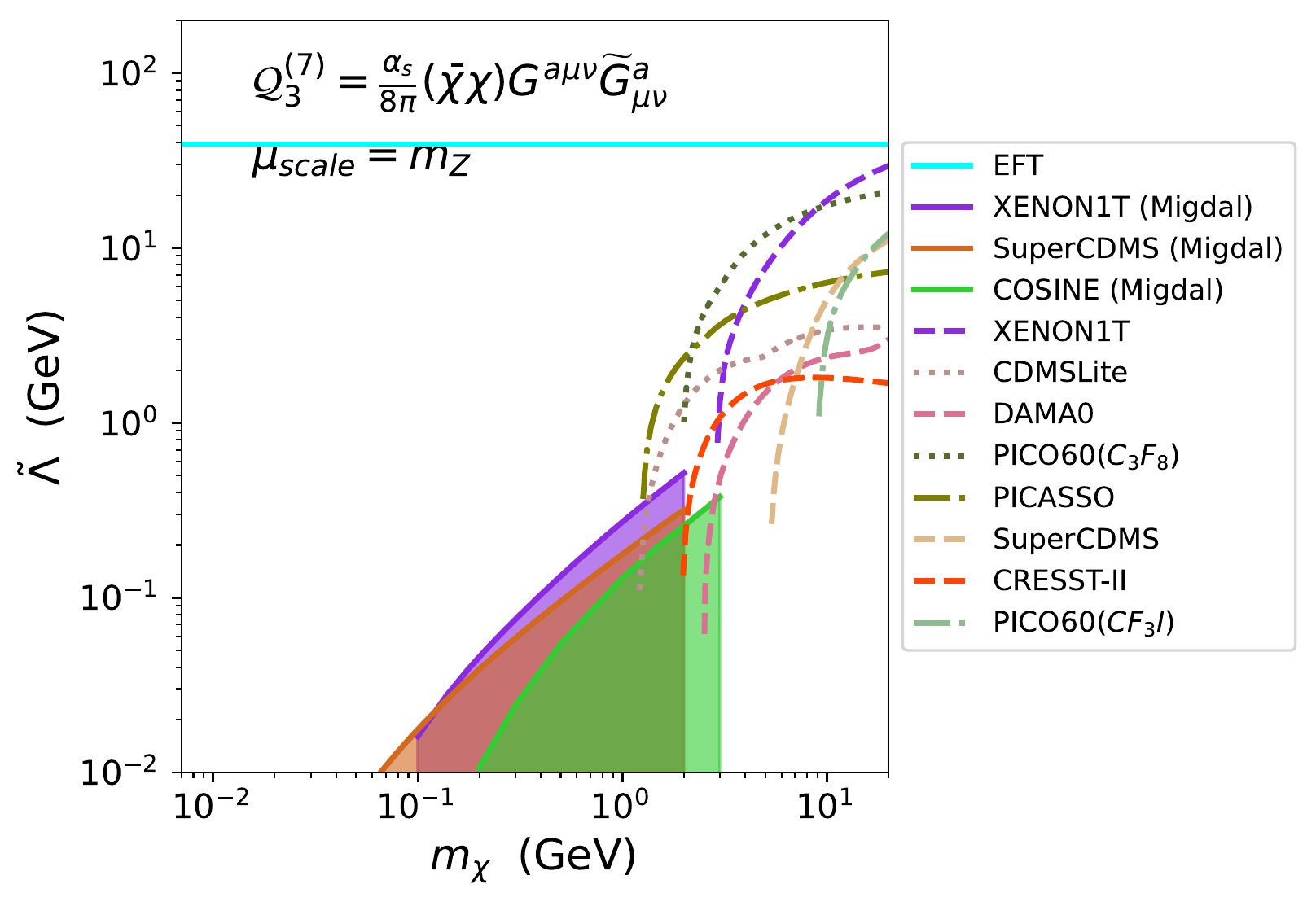}
  \includegraphics[width=0.49\textwidth]{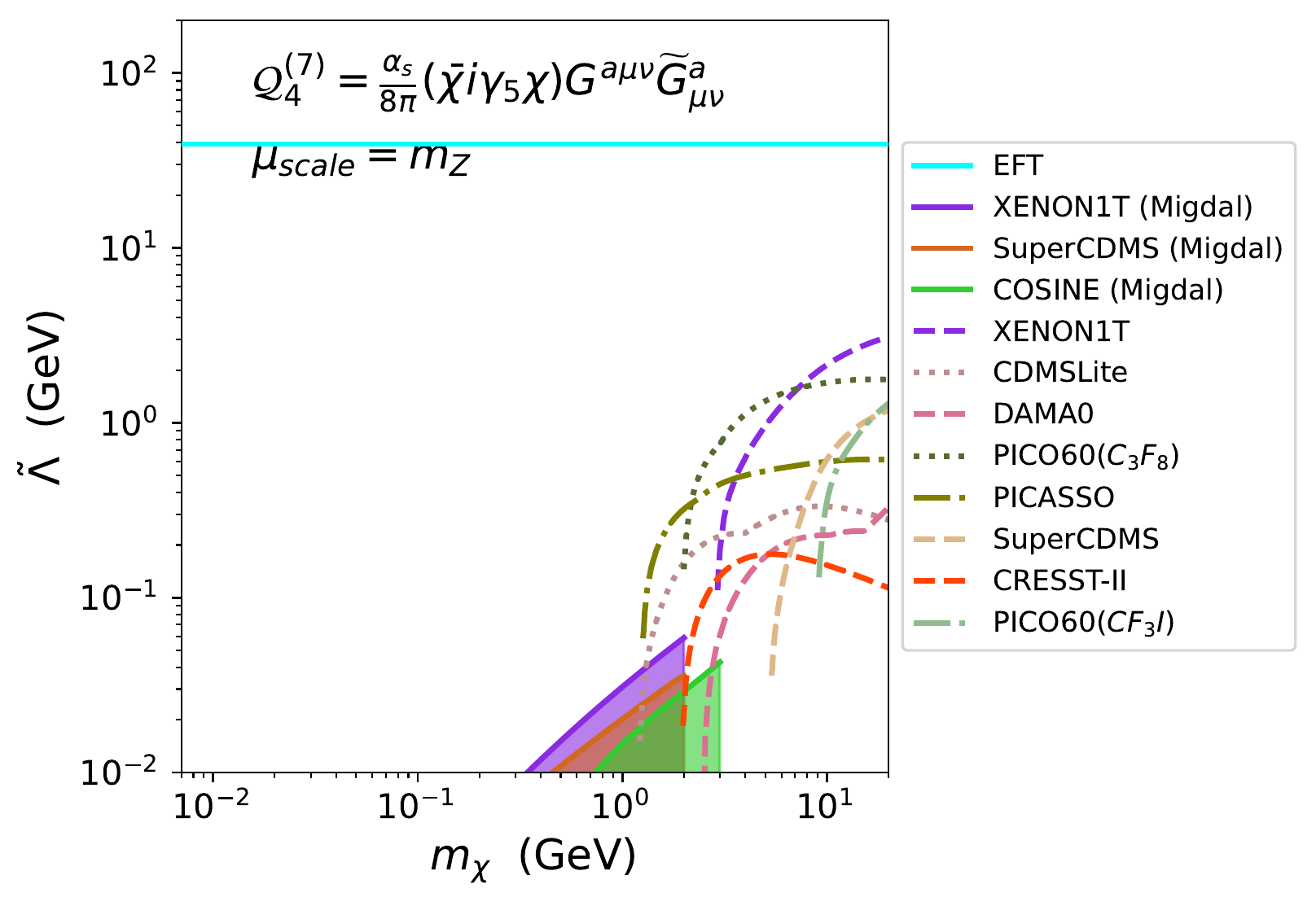}\\
   \includegraphics[width=0.49\textwidth]{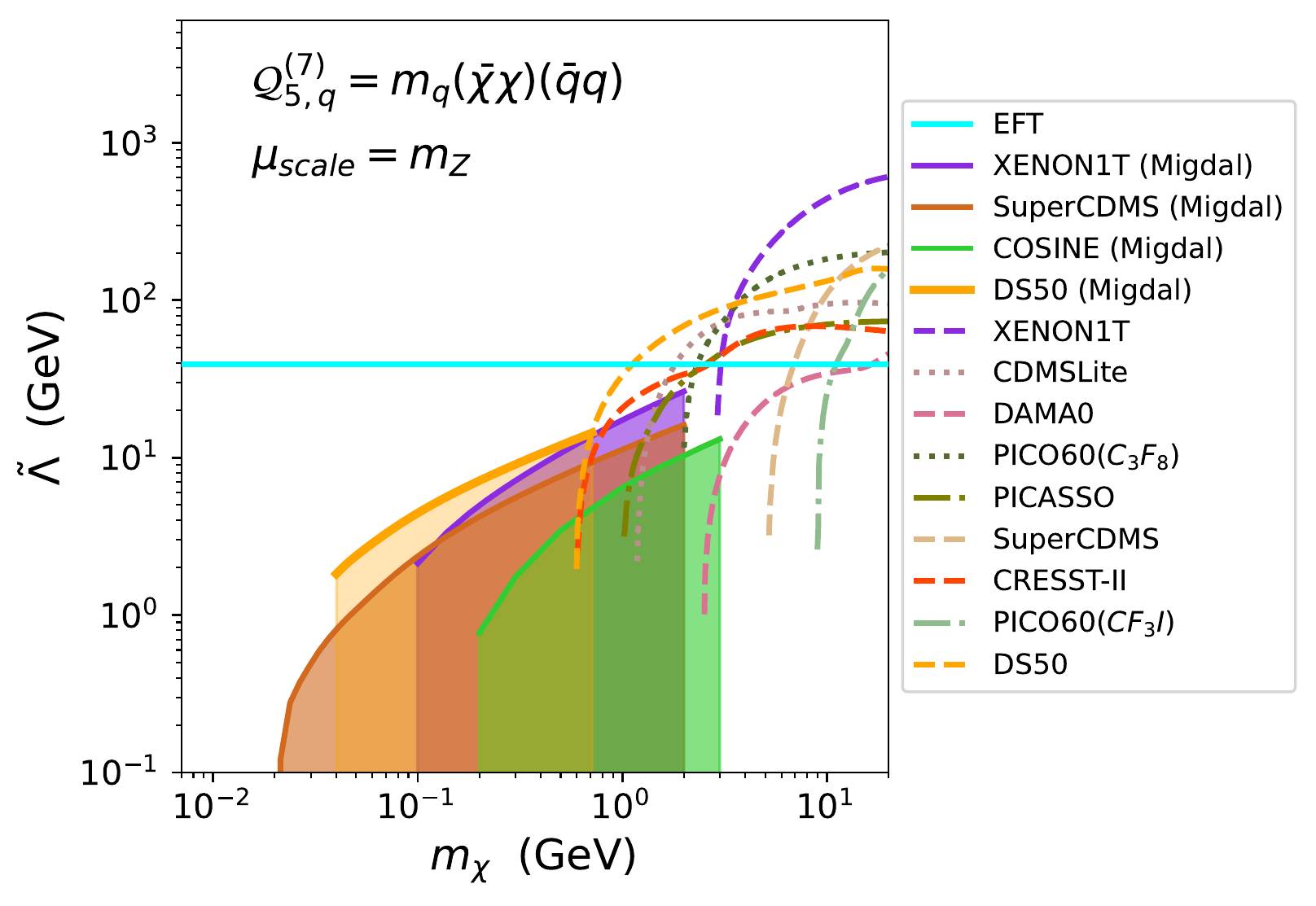}
  \includegraphics[width=0.49\textwidth]{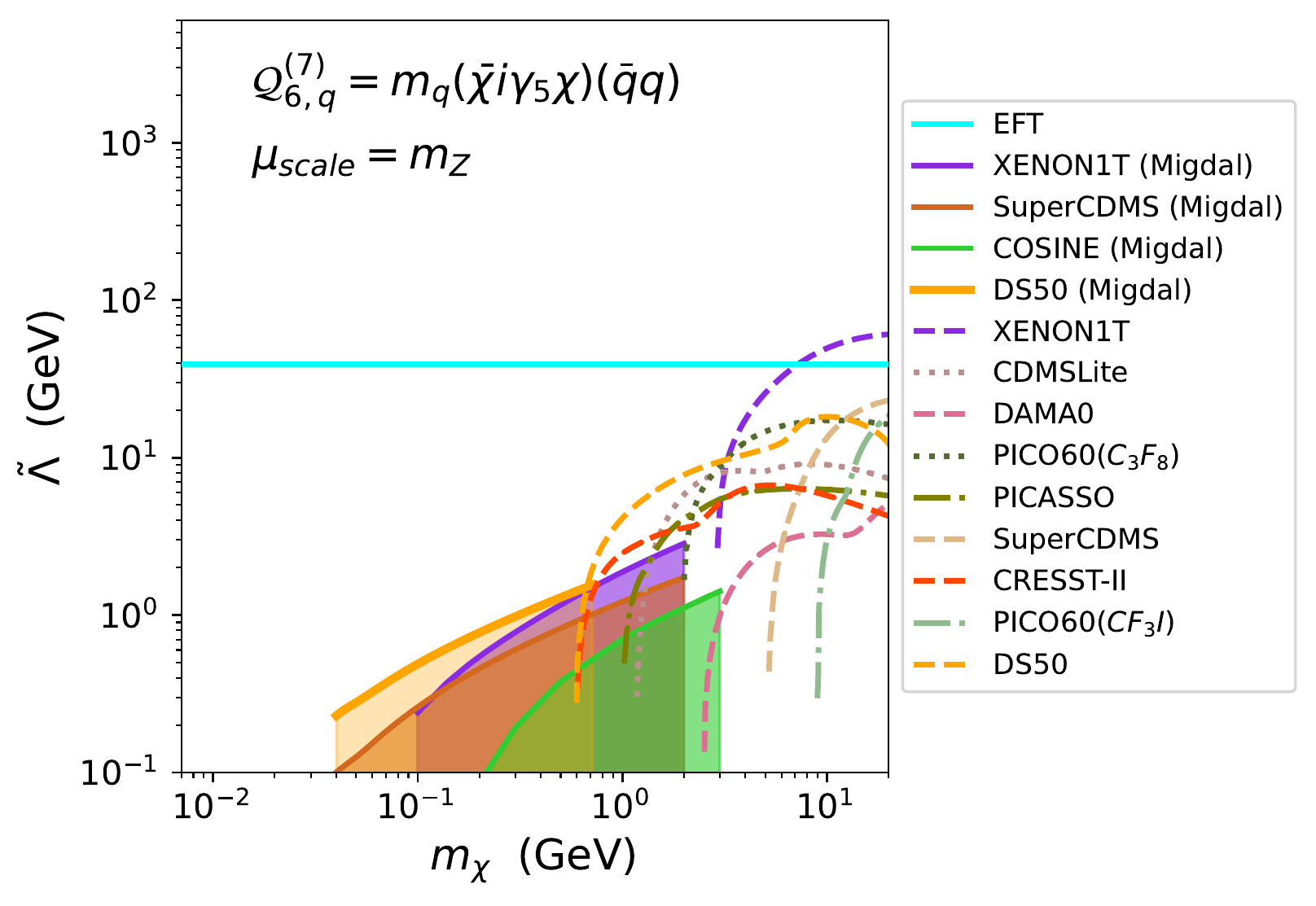}
  \end{center}
  \caption{The same as Fig.~\ref{fig:dim-5} for the ${\cal Q}_{1}^{(7)}-{\cal Q}_{6}^{(7)}$ operators.\label{fig:dim-7_1}}
\end{figure}

\begin{figure}[ht]
\begin{center}
  \includegraphics[width=0.49\textwidth]{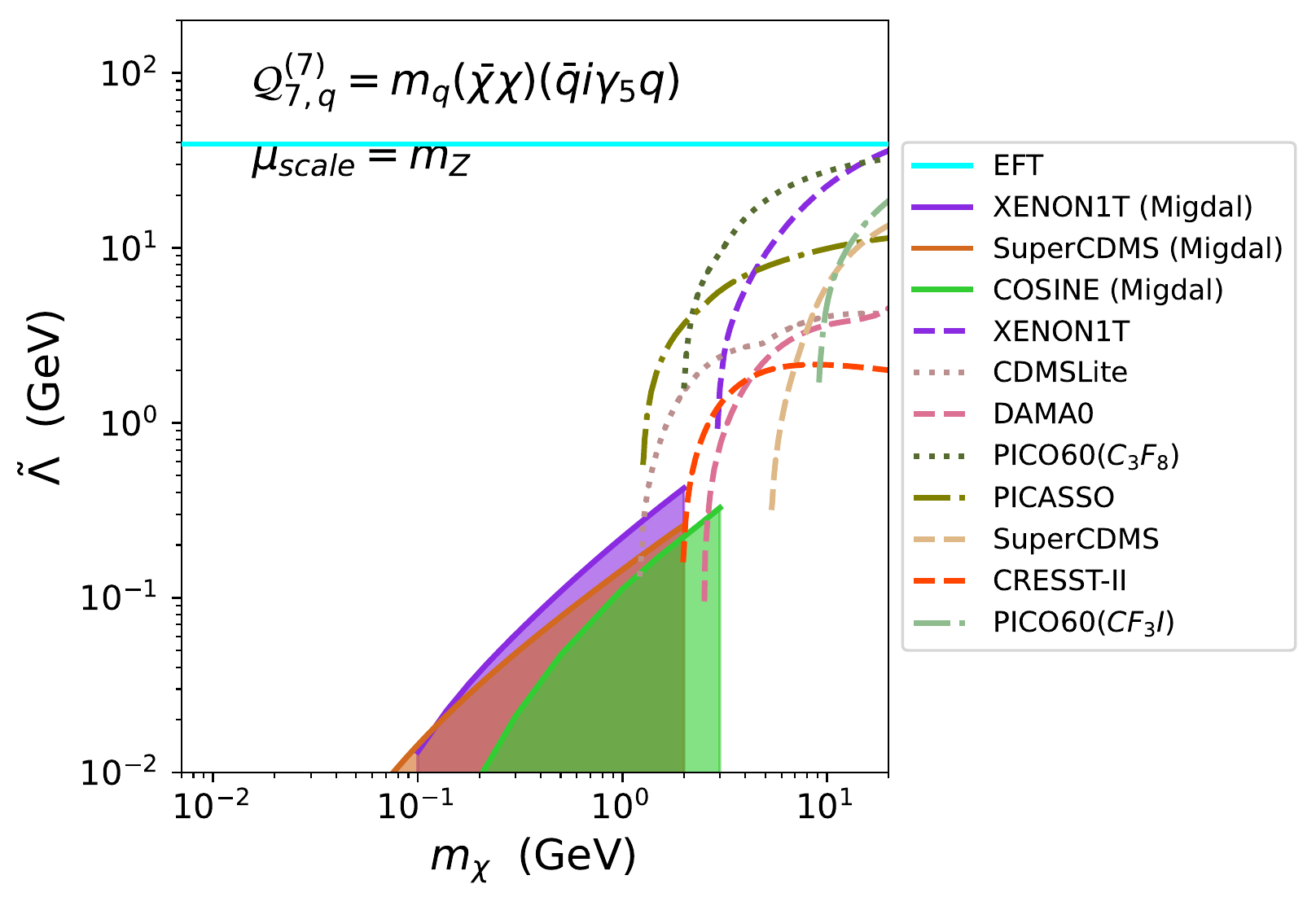}
  \includegraphics[width=0.49\textwidth]{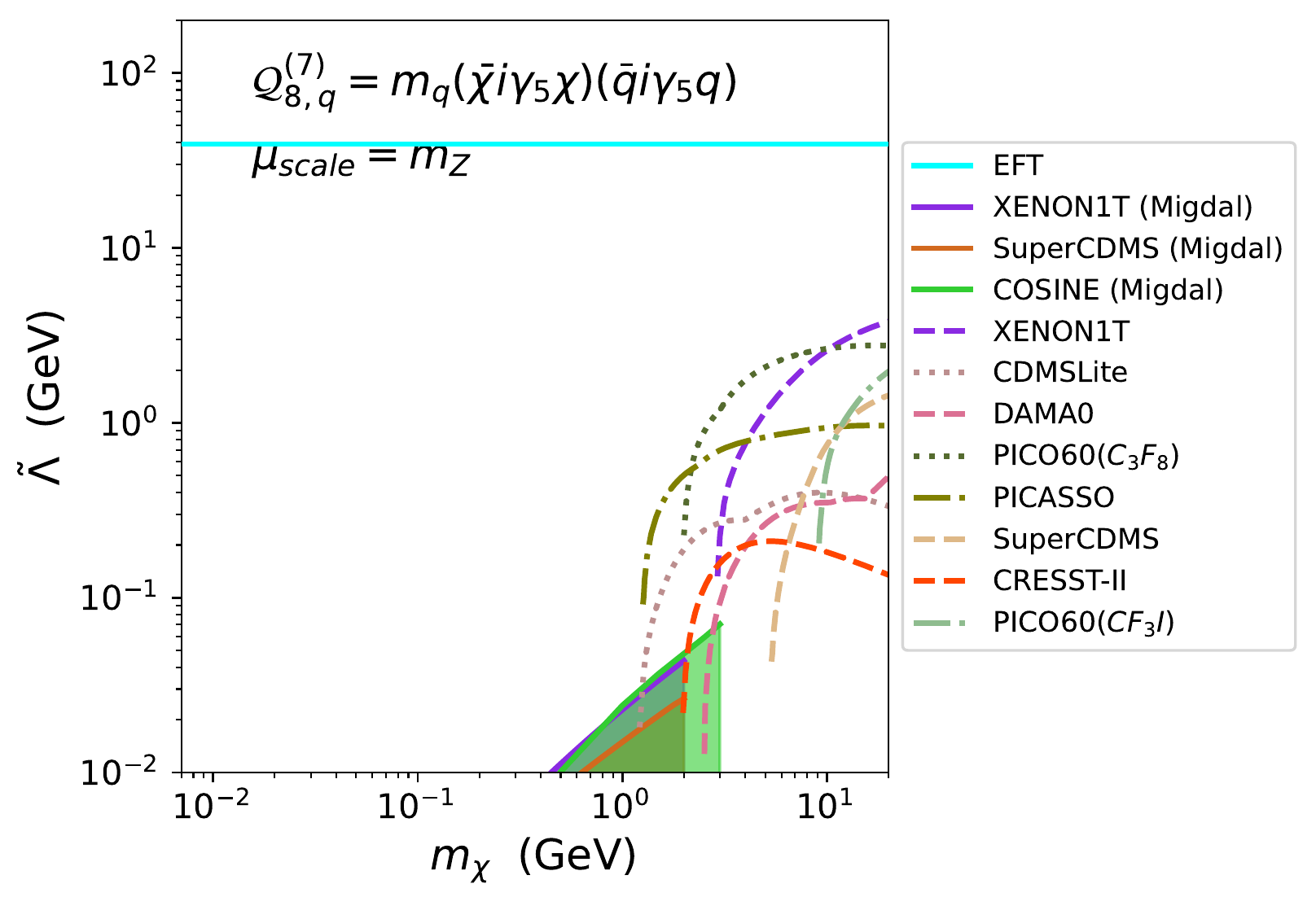}\\
    \includegraphics[width=0.49\textwidth]{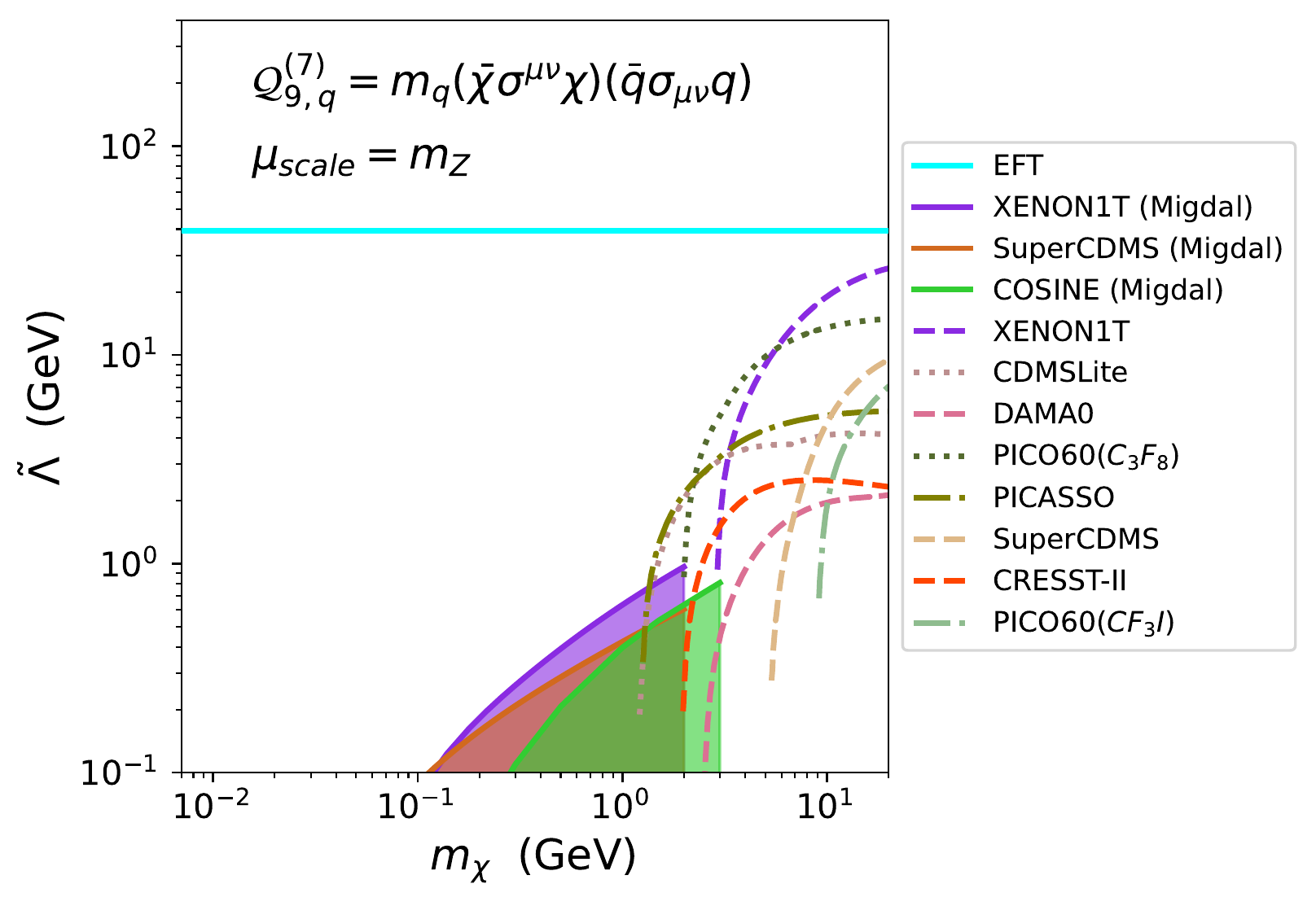}
  \includegraphics[width=0.49\textwidth]{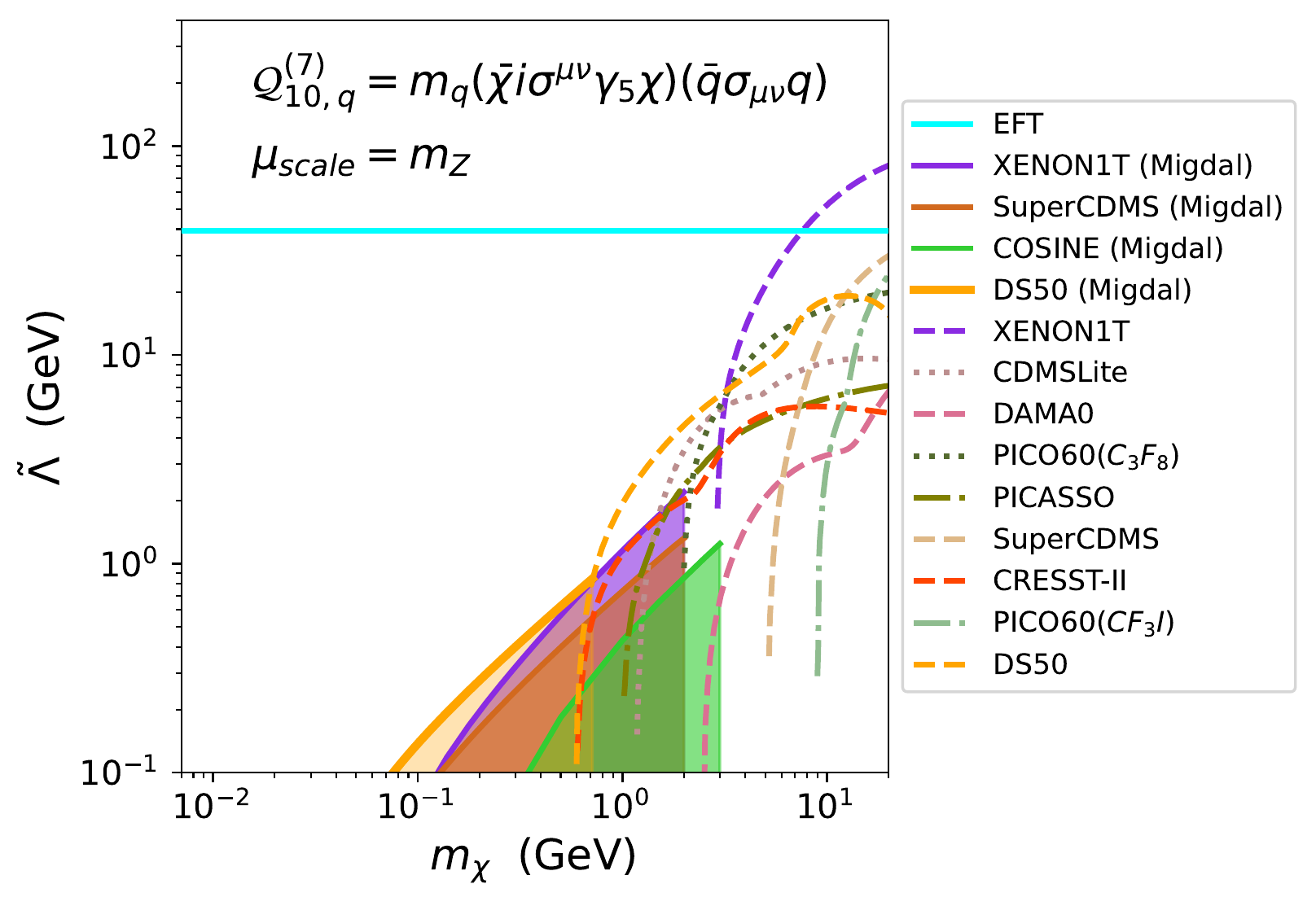}
\end{center}
\caption{The same as Fig.~\ref{fig:dim-5} for the ${\cal Q}_{7}^{(7)}-{\cal Q}_{10}^{(7)}$ operators.\label{fig:dim-7_2}}
\end{figure}

The results of our analysis are shown in Figs.~\ref{fig:dim-5}, \ref{fig:dim-6}, \ref{fig:dim-7_1} and \ref{fig:dim-7_2}. In particular adopting the same approach of Ref.~\cite{relativistic_eft_sogang} we fix 
the couplings ${\cal C}^{(d)}_{a,q}$ to a value common to all quarks and
show the constraint on each of the Wilson coefficient ${\cal C}^{(d)}_{a,q}$ and  ${\cal C}^{(d)}_{b}$ of Eq.~(\ref{eq:eff_L}) in terms of a 90\%--C.L. lower bound on the effective scale $\tilde \Lambda$ according to
the parameterization of Eq.(\ref{eq:lambda_tilde}). In the same plots we include for completeness the corresponding constraints from the elastic recoil analysis taken from~\cite{relativistic_eft_sogang}.

For XENON1T~\cite{XENON_migdal} we assume a 22 tonne-day exposure and $0.186\le E_{EM}\le 3.8$ keVee, with
49 WIMP candidate events corresponding at 90$\%$ C.L. to 61 observed events and an expected background of 23.4 events. 

For COSINE-100, we consider the first energy bin $1\le E_{EM}\le 1.25$ keVee, with an effective exposure of 97.7 kg--year. In the same bin from the upper panel of Fig. 4 of Ref.~\cite{cosine_migdal} the measured count--rate is $\simeq$ 20000 events, while from the lower panel the data exceed the estimated background by $\simeq$ 10\% at 90\% C.L. Using this piece of information we obtain $\simeq$ 2000 WIMP candidate events, which reproduce the published exclusion plot. 

Recently, the SuperCDMS~\cite{supercdms_migdal} collaboration has also published a dedicated Migdal analysis. Two separate sets of data are considered, corresponding to exposures of 18.8 kg-days and 17.5 kg-days. For both sets we consider a single energy bin 0.07$\le E_{EM}\le$2 keVee, with 208 and 193 WIMP candidate events~\cite{supercdms_data}, respectively. In our plots we show the most constraining bound between the two. The efficiency and resolution of SuperCDMS are taken from~\cite{supercdms_data}. 

The profile--likelihood analysis used by  DarkSide-50~\cite{ds50_migdal} is difficult to reproduce, but we notice that the Migdal energy spectrum $dR/dE_{EM}$ is fixed by the ionization probabilities $p^c_{q_e}$ and is the same for all interactions. As a consequence we directly use the normalization of the exclusion plot in~\cite{ds50_migdal}, obtained for a standard spin--independent interaction, to estimate the upper bound on the WIMP candidate events for all other interactions. In order to reproduce the exclusion limit of Fig.~3 in Ref.~\cite{ds50_migdal} we adopt energy bin close to threshold,  $0.083\le E_{EM}\le 0.106$ keVee, where we estimate 20 events, for an exposure of $\simeq$ 12.5 tonne-day. 

For the sake of comparison, in Figs.~\ref{fig:dim-5}, \ref{fig:dim-6}, \ref{fig:dim-7_1} and \ref{fig:dim-7_2}  we also provide the results obtained in Ref.~\cite{relativistic_eft_sogang} using elastic scattering and that extend down to $m_\chi\simeq$ 600 MeV. Indeed, the use of the Migdal effect  allows to extend the sensitivity of DD searches to WIMP masses that are significantly lower compared to the analysis in~\cite{relativistic_eft_sogang}, and that can reach down to $m_\chi\simeq$ 20 MeV. In all the plots the experiment that is sensitive to the lowest WIMP masses is SuperCDMS, with the lowest energy threshold at 70 eV. On the other hand
COSINE-100 has the higher threshold at 1 keVee and is never competitive in the determination of the constraints, with the exception of ${\cal Q}_{8,q}^{(7)}$. 

Figs.~\ref{fig:dim-5}, \ref{fig:dim-6}, \ref{fig:dim-7_1} and \ref{fig:dim-7_2} can be roughly divided in two classes: in the case of operators ${\cal Q}^{(5)}_{1,q}$, ${\cal Q}^{(5)}_{2,q}$, ${\cal Q}^{(6)}_{1,q}$, ${\cal Q}^{(6)}_{2,q}$, ${\cal Q}^{(7)}_{1}$, ${\cal Q}^{(7)}_{2}$, ${\cal Q}^{(7)}_{5,q}$, ${\cal Q}^{(7)}_{6,q}$ and ${\cal Q}^{(7)}_{10,q}$ the scattering cross section is driven 
in the non-relativistic limit by the $W_M$ nuclear response function, which corresponds to a coherent spin--independent interaction. In this case for $m_\chi\gsim$ 40 MeV DarkSide-50 is the most constraining experiment, while the constraint from XENON1T starts at $m_\chi\simeq$ 100 MeV and reaches the same sensitivity of DarkSide-50 at $m_\chi\simeq$ 1 GeV. 

A second class of exclusion plots is represented by the operators ${\cal Q}^{(6)}_{3,q}$, ${\cal Q}^{(6)}_{4,q}$, ${\cal Q}^{(7)}_{3}$, ${\cal Q}^{(7)}_{4}$, ${\cal Q}^{(7)}_{7,q}$, ${\cal Q}^{(7)}_{8,q}$, and ${\cal Q}^{(7)}_{9,q}$, for which, instead, in the non-relativistic limit the scattering cross section is driven by a nuclear response function of the spin--dependent type (either $\Sigma^{\prime\prime}$, $\Sigma^{\prime}$ or both). In this case the DarkSide-50 bound is not present because Argon ($^{40}$Ar) has no spin, so that for $m_\chi\gsim$ 100 MeV XENON1T is the most constraining bound. 
One exception is represented by the ${\cal Q}^{(6)}_{3,q}$ operator that develops a ${\cal Q}^{(6)}_{1,q}$ component driven by the $W_M$ nuclear response function in the running from $m_Z$ to the nucleon scale~\cite{deramo_2014,deramo_2016}. This leads to a non-vanishing bound from DarkSide-50 for $m_\chi\gsim$ 40 MeV, that turns out to be at the same level of XENON1T.
Some of the limits shown in
Figs.~\ref{fig:dim-5}, \ref{fig:dim-6}, \ref{fig:dim-7_1} and \ref{fig:dim-7_2} may be so weak
that they put bounds on values of the $\tilde{\Lambda}$ scale which
are inconsistent with the validity of the effective theory. In such
case one can simply conclude that the present experimental sensitivity
of direct detection experiments is not able to put bounds on the
corresponding effective operator. A possible (but not unique) criterion for the validity of the
EFT is the same that we adopted in Ref.~\cite{relativistic_eft_sogang}. In this case the scale $\tilde{\Lambda}$ 
is interpreted in terms of a propagator $g^2/M_{*}^2$ with $g<\sqrt{4\pi}$ and $M_{*}>\mu_{scale}$,
with $\mu_{scale}$ = $m_Z$ the scale were we fixed the boundary conditions of the EFT. 
This is straightforward for dimension--6
operators, while in the case of operators whose effective coupling has
dimension different from -2 only matching the EFT with the full theory
would allow to draw robust conclusions. In particular, in this case
$\tilde{\Lambda}$ can be interpreted in terms of the same propagator
times the appropriate power of a typical scale of the problem
$\mu_{scale}^{\prime}$, which depends on the ultraviolet completion of
the EFT. For instance, in the operator ${\cal Q}_{5,q}^{(7)}$=$m_q
(\bar \chi \chi)( \bar q q)$ the quark mass may originate from a
Yukawa coupling, so $\mu_{scale}^{\prime}$ corresponds to the Electroweak vacuum
expectation value.  To fix an order of magnitude we
choose to fix $\mu_{scale}^{\prime}$ = $\mu_{scale}$, so that the
bound $\tilde{\Lambda}>\mu_{scale}/(4\pi)^{1/(d-4)}$ can be
derived. Such limit is shown as a horizontal solid line in
Figs.~\ref{fig:dim-5}, \ref{fig:dim-6}, \ref{fig:dim-7_1} and \ref{fig:dim-7_2}. In particular,
for models ${\cal Q}^{(6)}_{3}$, ${\cal Q}^{(6)}_{4}$ in Fig.~\ref{fig:dim-6} and for all the dimension--7 operators of Figs.~\ref{fig:dim-7_1} and \ref{fig:dim-7_2} the extension at low WIMP masses of the exclusion plot on the $\tilde{\Lambda}$ scale obtained with our Migdal effect analysis lies below such horizontal line. This may imply that the sensitivities of the present direct detection experiments optimized to search for the Migdal effect is not sufficient to put
meaningful bounds at low WIMP masses. However we stress again that this can only be
assessed when a specific ultraviolet completion of the effective
theory is assumed.
\section{Conclusion}
\label{sec:conclusion}

In a previous analysis~\cite{relativistic_eft_sogang} we studied the direct detection bounds from elastic WIMP--nucleus scattering  to operators up to dimension 7 of the relativistic effective field theory describing WIMP interactions with quarks and gluons. Such bounds reached a WIMP mass $m_\chi\gsim$ 600 MeV. In the present letter we have used the inelastic Migdal effect, where the recoiling nucleus is ionized, to extend such bounds to lower WIMP masses. In particular, analyzing the data of  XENON1T, SuperCDMS, COSINE-100, and DarkSide-50 we find that the bounds can reach down to a WIMP mass $\simeq$20 MeV. In the case of higher--dimension operators the energy scale of the ensuing constraints may be inconsistent with the validity of the effective theory.

\section*{Acknowledgements} 
The work of G.T. was supported by the Collaborative Research Center SFB1258. The work of S.K. and S.S. was supported by the National Research Foundation of Korea (NRF) funded by the Ministry of Education through the Center for Quantum Space Time (CQUeST) with grant number 2020R1A6A1A03047877 and by the Ministry of Science and ICT with grant number 2021R1F1A1057119. G.T is thankful to Alejandro Ibarra and 
Maximilian Gapp for helpful discussions.
%


\bibliographystyle{elsarticle-num}


\end{document}